\pdfoutput=1
\documentclass[english]{aa}
\usepackage[T1]{fontenc}
\usepackage{array}
\usepackage{float}
\usepackage{units}
\usepackage{url}
\usepackage{amstext}
\usepackage{graphicx}
\usepackage{wasysym}

\makeatletter

\providecommand{\tabularnewline}{\\}

\usepackage{graphicx}
\usepackage[varg]{txfonts}
\usepackage{gensymb}

\usepackage{natbib,twoopt}
\usepackage[breaklinks=true]{hyperref} 
\hypersetup{colorlinks=true,citecolor=blue,linkcolor=blue,filecolor=black,runcolor=black,urlcolor=blue}
\bibpunct{(}{)}{;}{a}{}{,} 

\defcitealias{2015A&A...575A..57F}{Paper I}

\renewcommand\[{\begin{equation}}
\renewcommand\]{\end{equation}}

\usepackage{siunitx,booktabs}
\makeatother

\usepackage{babel}
\begin{document}

\title{Short-term evolution and coexistence of spots, plages and flare activity
on LQ Hydrae\thanks{Based on data obtained with the STELLA robotic telescope in Tenerife,
an AIP facility jointly operated by AIP and IAC, and the Vienna-Potsdam
Automatic Photoelectric Telescopes at Fairborn Observatory in Arizona,
operated by AIP.}}

\author{M. Flores Soriano \and  K. G. Strassmeier}
\authorrunning{Flores Soriano \and  Strassmeier}

\institute{Leibniz-Institut f\"ur Astrophysik Potsdam (AIP), An der Sternwarte
16, 14482 Potsdam, Germany \\
\email{{[}mflores;kstrassmeier{]}@aip.de}}

\date{Received ...; Accepted...}

\abstract{} {We aim to study the short-term evolution of the chromospheric
and photospheric activity of the young, single K2 dwarf LQ~Hya.} {Four
months of quasi-simultaneous spectroscopic and photometric observations
were used to study the variations of the photometric light curve,
the evolution of the chromospheric activity from the H$\alpha$ and
H$\beta$ lines, and the distribution of cool spots from Doppler maps.} {During
our observations one side of the star was more active than the other.
The equivalent width of the H$\alpha$ line from the least active
hemisphere increased from $\approx0.7$\,\AA\ at the beginning of
the observation to $1.0$\,\AA\ at the end. The basal emission of
the most active hemisphere remained roughly constant at $EW_{\mathrm{H\ensuremath{\alpha}}}\thickapprox1.0$\,\AA.
Intense flare activity was observed during the first twenty days,
where at least four different events were detected. The line asymmetries
of the H$\alpha$ line suggest that one of the flares could have produced
a mass ejection with a maximum projected speed of $\mathrm{70\thinspace km\thinspace s^{-1}}$.
The rotational modulation of the \emph{V}-band photometry showed clear
anti-correlation with the chromospheric activity. The difference in
brightness between the opposite hemispheres decreased from 0\fm16
to 0\fm09 in two months. Three spots gradually moving apart from
each other are dominating the photospheric Doppler maps. The comparison
between the maps and the H$\alpha$ line as the star rotates reveals
the spatial coexistence of chromospheric H$\alpha$ emission and photospheric
spots.} {Our results indicate that the active regions of LQ~Hya
can live for at least four months. The detected changes in the photometric
light curve and the spectroscopic Doppler images seem to be more a
consequence of the spatial redistribution of the active regions rather
than due to changes in their strength. Only one of the active regions
shows significant changes in its chromospheric emission.}

\maketitle

\section{Introduction}

LQ~Hya (HD 82558) is a young, single, fast rotating K2 dwarf with
high magnetic activity. It was first identified as a chromospherically
active star by the surveys of \citet{1981AJ.....86..553B} and \citet{1981ApJS...46..247H}.
\citet{1986ApJS...60..551F} observed the H$\alpha$ line completely
filled-in by emission and classified the star as a BY Draconis variable.
Age estimations suggest that the star could probably have just arrived
on the zero-age main sequence. From its fast rotation and high lithium
abundance, \citet{1986AJ.....92.1150F} concluded that LQ~Hya is
not older than 75\,Myr, \textquotedblleft at least as young as the
youngest Pleiades star\textquotedblright . The lack of persistent
variability in the radial velocity of LQ~Hya indicates that it is
a single star (see, e.g., \citealt{1986AJ.....92.1150F}; \citealt{2003MNRAS.345.1145D}).
Nonetheless, occasional rotational modulations produced by asymmetric
spot distributions have also been observed (\citealt[hereafter \citetalias{2015A&A...575A..57F}]{2015A&A...575A..57F}).

Because of its physical properties, LQ~Hya is commonly considered
a young solar-like star, and a good link between the very active and
fast rotating pre-main sequence stars and the more evolved stars from
the main sequence that have already undergone rotational momentum
loss. It has been the target of numerous photometric studies. The
rotational modulation of its light curve indicates that the star rotates
with a period of $P_{\mathrm{rot}}\thickapprox1.6$\,d. With data
available since the early 1980s, many authors have found signs of
periodic spot activity in three time ranges: a short cycle with a
period between 2 and 3\,yr (\citealt{2002AN....323..361O}; \citealt{2003A&A...409.1017M};
\citealt{2004A&A...417.1047K}; \citealt{2009A&A...501..703O}; \citealt{2016A&A...588A..38L}),
an intermediate cycle with a period in the range 6-7\,yr (\citealt{1993A&A...276..345J};
\citealt{1997A&AS..125...11S}; \citealt{2000A&A...356..643O}; \citealt{2002A&A...394..505B};
\citealt{2003A&A...409.1017M}; \citealt{2004A&A...417.1047K}), and
a longer cycle with a 11-18\,yr period (\citealt{2000A&A...356..643O};
\citealp{2002A&A...394..505B}; \citealt{2003A&A...409.1017M}; \citealt{2004A&A...417.1047K};
\citealt{2012A&A...542A..38L}; \citealt{2016A&A...588A..38L}). \citet{2002A&A...394..505B}
found that two active longitudes dominated the activity over 20 years
and reported an additional 5.2\,yr flip-flop cycle. On the contrary,
\citet{2012A&A...542A..38L} and \citet{2015A&A...577A.120O} have
not found any periodicity for this kind of events. Coherent active
longitudes were found by \citet{2016A&A...588A..38L} only between
2003 and 2008. The time scale of the changes in the light curve has
been reported to be of a few months by \citet{1993A&A...276..345J}
and 50.5\,d by \citet{2012A&A...542A..38L}.

The star is also known for being chromospherically active. \citet{1993A&A...268..671S},
\citet{2008A&A...481..229F}, \citet{2014AJ....147...38C}, and \citetalias{2015A&A...575A..57F}
found a clear anti-correlation between the rotational modulation of
the chromospheric emission and the photometric light curve. This points
out the spatial connection between photospheric dark spots, and chromospheric
plages and possibly microflare activity (see \citetalias{2015A&A...575A..57F}
for a characterization of the emission sources). In addition to this,
\citet{2014AJ....147...38C} also found that the chromospheric activity
gradually decreased from 2006 to 2012, which corresponds to a period
of constant increment in the average brightness of the star (\citealt{2016A&A...588A..38L}).

Strong flares have been reported in LQ~Hya in different wavelengths.
\citet{1990BAAS...22..857A} detected an UV flare, but curiously without
enhancement in the chromopheric lines. On the other hand, \citet{1999MNRAS.305...45M}
observed a flare in both optical and UV wavelengths, but with the
optical flare apparently delayed. They found that the chromospheric
lines reached their maximum intensity $\approx55$\,min after the
impulsive phase. \citet{2001A&A...371..973C} observed an X-ray flare
and derived a loop semi-length of $\approx1.5$ stellar radii. Frequent
low-level flare activity was also reported by \citet{1998ASPC..154.1560S}
and in \citetalias{2015A&A...575A..57F}. 

LQ Hya has been regularly analyzed with surface mapping techniques
since the early 1990s (\citealt{1992ASPC...26..255S}; \citealt{1993A&A...268..671S};
\citealt{1994ASPC...64..661S}; \citealt{1998A&A...336..972R}). Nearly
annual magnetic and brightness maps were computed by \citet{1999MNRAS.302..457D}
and \citet{2003MNRAS.345.1145D} for the period 1991 - 2001. Their
brightness images usually show a polar spot and some low latitude
spots, but with clear changes in the configuration from one year to
the next. \citet{2001ASPC..223.1207B} computed temperature maps from
1993 to 1999 with a typical time interval between images of about
four months. They reported a rapid evolution of the active regions,
on a time scale of months. \citet{2004A&A...417.1047K} reconstructed
28 maps from 35 consecutive stellar rotations and saw rapid spot evolution
on time scales from 1 day to 20 days. They interpreted the fast changes
in the spot configuration as due to magnetic reconnection rather than
due to spot migration or the emergence of new flux tubes. More recently,
\citet{2015A&A...581A..69C} computed seven maps for the observing
seasons 1998 \textendash{} 2002. They addressed a chaotic spot activity
but apparently concentrated at two latitudes.

In spite of all this progress, the short term evolution of the magnetic
activity of LQ~Hya remains puzzling. Studies based on spectral data
are usually either from observing campaigns that are not long enough
to show any significant change, or from interseasonal data that are
too sparse to follow the evolution of individual magnetic regions.
In \citetalias{2015A&A...575A..57F} we analyzed a time series of
199 LQ~Hya spectra and quasi-simultaneous photometry to study the
rotational modulation of chromospheric and photospheric parameters
and their connection with the variability of the \ion{Li}{I} 6708\,\AA\
line. We found that during the four months of nearly continuous observations
the star presented a relatively stable activity configuration, where
one side of the star was clearly more active than the other. In this
paper we use the superior spectral time coverage of those observations
to study the temporal evolution of the magnetic activity of LQ~Hya
in a previously unexplored timescale. Sect. \ref{sec:2Observations}
describes our data and observations. The temporal evolution of the
chromospheric emission at different phases is presented in Sect. \ref{sec:3Ha_Emission}.
The photospheric variability is analyzed in Sect. \ref{sec:Evo_fotosfera}
with photometry and by means of Doppler imaging. Sect. \ref{sec:Putting-things-together}
combines our results and explores the connection between the photospheric
spots and the chromospheric activity. Finally, Sect. \ref{sec:Summary-and-conclusions}
summarizes and concludes the paper.

\section{Observations and data reduction \label{sec:2Observations}}

\subsection{STELLA/SES spectroscopy}

For this work we used the same spectroscopic data as in \citetalias{2015A&A...575A..57F}.
They are a time series of 199 echelle spectra taken between December
2011 and April 2012 with the STELLA Echelle Spectrograph (SES) at
the robotic 1.2 m STELLA-I telescope at the Observatorio del Teide
in Tenerife, Spain \citep{2010AdAst2010E..19S}. The integration time
was set to 3600\,s, providing signal-to-noise ratios (S/N) between
40 and 120. During the observations the spectrograph covered the wavelength
range from 476 to 764\,nm with a resolving power of $R=30\,000$.
The SES spectrograph normally operates with a fixed spectral range
of 388\textendash 882\,nm and $R=55\,000$. However, our observations
were made during an interruption in the ongoing spectrograph upgrade.
At that time the collimator was slightly misaligned and the old camera
out of focus, which caused the shorter spectral range and the significantly
lower spectral resolution. For more details about the instruments
and data reduction we refer to \citetalias{2015A&A...575A..57F} and
references therein. 

\subsection{Amadeus APT photometry}

Quasi-simultaneously to the STELLA observations, phase-resolved $V$-band
photometry was obtained with the 0.75\,m Potsdam-APT {\sl Amadeus}
(T7) telescope at Fairborn Observatory in southern Arizona \citep{1997PASP..109..697S}.
They are basically the same data as in \citetalias{2015A&A...575A..57F},
but we have added new values to improve the time and phase coverage
of our evolution analysis. The total amount of measurements is now
1144. All measurements were made differentially with respect to HD~82447
and HD~82508 as the comparison and check star, respectively. The
integration time was 10\,s. Data taken before 2012-02-20 are affected
by instrumental problems. They are substantially more disperse and
have larger error bars. Most of these values have been removed and
only those of apparently better quality are kept. For further details
about the observing procedure and the APT data reduction, we refer
to \citet{2001AN....322..325G}. The uncertainties were calculated
as indicated in \citet{1986IAPPP..25...32H}.

\section{Evolution of the H$\alpha$ and H$\beta$ chromospheric emission
\label{sec:3Ha_Emission}}

In \citetalias{2015A&A...575A..57F} we studied the rotational modulation
of the chromospheric activity of LQ~Hya from the emission of the
Balmer H$\alpha$ and H$\beta$ lines. We evaluated the equivalent
widths (EWs) of the net emission profiles by integrating the spectra
resulting from the subtraction of a non-active template. We found
that the chromospheric emission in these lines can be divided into
two different types of contributions. The first contribution is a
relatively stable emission with H$\alpha$ EWs in the range $0.7-1.2$\,\AA\
that we attributed to a combination of plages and repeated low-intensity
flares. It is rotationally modulated and presents maximum values when
the most spotted side of the star is in view. The other contribution
consists in more energetic episodes ($EW_{\mathrm{H\alpha}}\apprge1.2$\,\AA)
that correspond to sporadic flares of moderate intensity, occurring
predominantly also on the most spotted side of the star. In this section
we will discuss the temporal evolution of these chromospheric emissions.
As the H$\alpha$ and H$\beta$ lines present a similar behavior,
we will focus our analysis on the variations of the H$\alpha$ line.
The numerical values of the EWs of both lines, their associated radiative
flux, and the Balmer decrements $F_{\mathrm{H\ensuremath{\alpha}}}/F_{\mathrm{H\ensuremath{\beta}}}$
are available at the online material of \citetalias{2015A&A...575A..57F}\footnote{Available at the CDS via anonymous ftp to\\
\url{cdsarc.u-strasbg.fr (130.79.128.5)} or via\\
\url{http://cdsarc.u-strasbg.fr/viz-bin/qcat?J/A+A/575/A57}}.

\subsection{Sporadic flare emission}

\begin{figure}
\includegraphics[width=1\columnwidth]{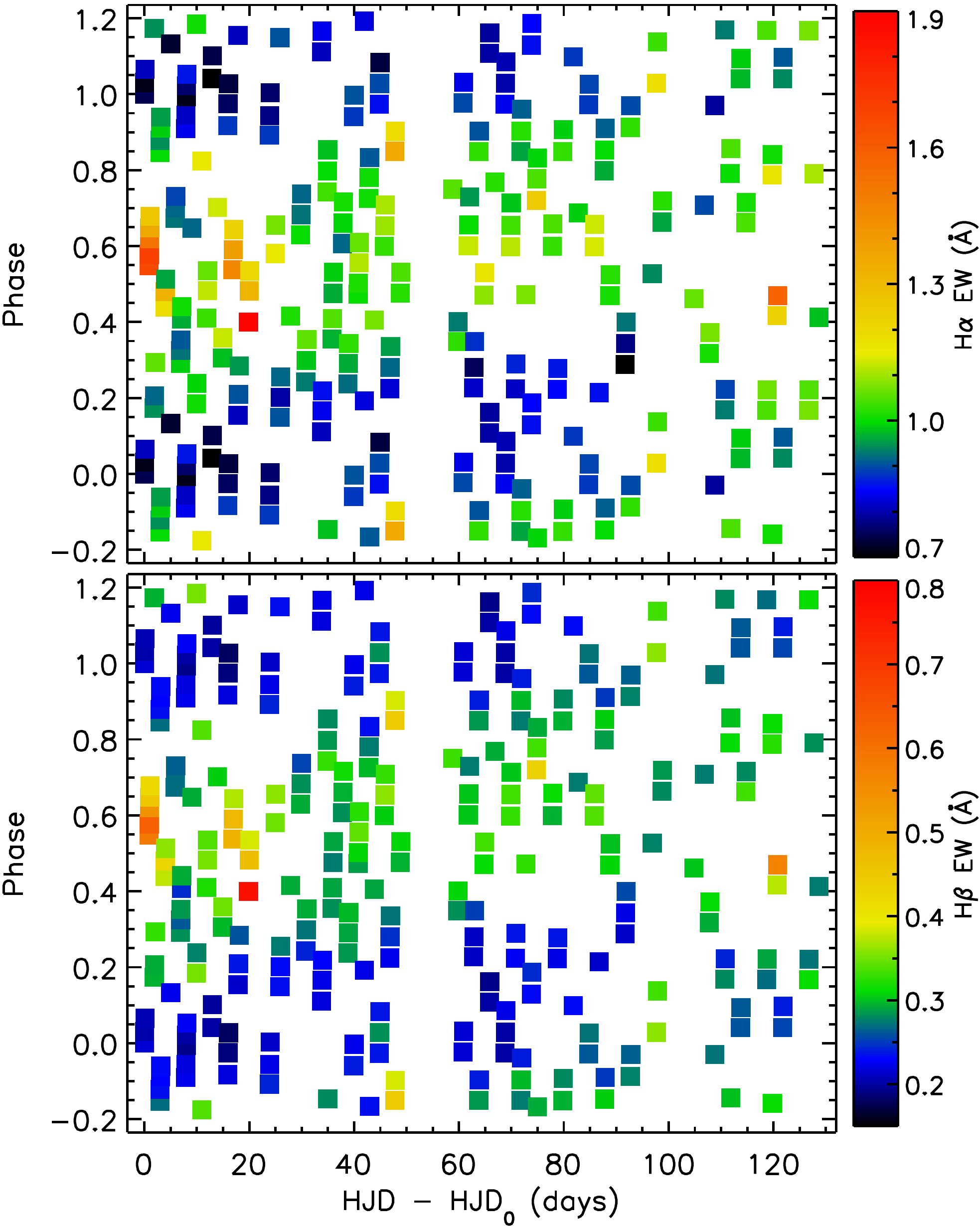}

\caption{Temporal evolution of the equivalent width of the H$\alpha$ and H$\beta$
emission profiles (top and bottom panels, respectively) at different
rotational phases. Blue-green points correspond to measurements of
the basal emission. Orange-red values correspond to sporadic flares.
\label{fig:TimePhaseEW}}
\end{figure}

The top panel of Figure \ref{fig:TimePhaseEW} shows the EW of the
H$\alpha$ line for each one of the 199 LQ~Hya spectra at different
times and phases. As in \citetalias{2015A&A...575A..57F}, we phased
the observations with the ephemeris $\mathrm{HJD}=\mathrm{HJD_{0}}+1.60066\cdot E$,
where the zero epoch $\mathrm{HJD_{0}}=2455919.67$ corresponds to
the first spectrum, and the rotation period of the star is from \citet{2004A&A...417.1047K}.
Times are given relative to $\mathrm{HJD_{0}}$. The color table of
the figure was prepared to represent the basal emission with blue-green
colors and the sporadic flares with orange-red colors. The basic criterion
that we used is that the EW of the H$\alpha$ flare contribution should
be clearly larger than the EW of the neighboring basal contribution.
A clear and unambiguous distinction between the two sources was nevertheless
not always possible. To prevent the misclassification of optically-thin
material seen off-limb as flares, we also required that the Balmer
decrement $F_{\mathrm{H\ensuremath{\alpha}}}/F_{\mathrm{H\ensuremath{\beta}}}$
should not exceed the neighboring values. However, this approach is
maybe too simple to account for some of the H$\alpha$ lines with
EWs near 1.2\,\AA. In principle, they could correspond, for example,
to a slightly higher than usual basal emission due to enhanced microflare
or network activity, the early or late stages of a more powerful flare,
or an intense event but with small projected area. This makes difficult
to be sure about the exact amount of flares detected, but using EWs
of 1.2\,\AA\ and 1.15\,\AA\ as threshold values we believe to
have seen between 7 and 12 flare events, respectively. The H$\beta$
line shows only 7 events whose emission is clearly above the basal
contribution (bottom panel of Fig.~\ref{fig:TimePhaseEW}).

It appears that the flare activity was more intense at the beginning
of our observing campaign than at later stages. Almost as many flares
were detected during the first twenty days as during the following
three and a half months. They were also more intense and occurred
concentrated between phases 0.4 and 0.7. As a consequence of the fast
rotation of the star and the given sampling of the observations, none
of the detected flares was observed in its full extension. The total
duration of the events is therefore unknown, but in some cases it
is longer than five hours. A more detailed description of the flare
with the best time coverage and its post-flare phase is presented
in Sect.~\ref{subsec:-ME?}. 

\subsection{Basal emission}

It is well known that LQ~Hya exhibits cool spots that correlate well
with the chromospheric emission (\citealt{1993A&A...268..671S}; \citealt{2008A&A...481..229F};
\citealt{2014AJ....147...38C}; \citetalias{2015A&A...575A..57F}).
In \citetalias{2015A&A...575A..57F} we found that the hemisphere
of the star centered at phase 0.6 had during our observations a higher
concentration of spots and enhanced chromospheric activity than the
other side of the star. The difference in the basal emission of these
two hemispheres is evident from Fig. \ref{fig:TimePhaseEW}, where
they appear as a green and as a blue band, respectively. The basal
emission of the most active hemisphere remained roughly constant during
the four months of observations with $EW_{\mathrm{H\ensuremath{\alpha}}}\thickapprox1.0$\,\AA.
We also see that the active regions seem to be moving toward higher
phases, surely because their rotation period is slightly longer than
the rotation period that we have used to phase our data.

The less active hemisphere of the star shows, on the other hand, a
clear temporal evolution. We detect the lowest level of H$\alpha$
emission at the beginning of our observation and near phase 0.0. The
EW of the subtracted H$\alpha$ line takes then values of $\approx0.7$\,\AA,
which is still a relatively high level of activity. LQ Hya H$\alpha$
lines with this emission have their core almost at the continuum level
(see Fig. 2. in \citetalias{2015A&A...575A..57F}). The chromospheric
activity from that hemisphere undergoes then a slow but constant rise
in its emission. The H$\alpha$ equivalent width takes values of $\approx0.8$\,\AA\
at the middle of the observations, and between 0.9 and 1.1\,\AA\
during the last two weeks, meaning that the less active side of the
star reached at the end of the observations values that until then
had been detected only on the opposite hemisphere. Nonetheless, its
average H$\alpha$ emission was still slightly below that of the opposite
side. The H$\beta$ line, on the other hand, shows a more pronounced
rotational modulation when the observations finished.

\subsection{H$\alpha$ line asymmetries: possible mass ejection \label{subsec:-ME?}}

\begin{figure*}
\includegraphics[width=1\textwidth]{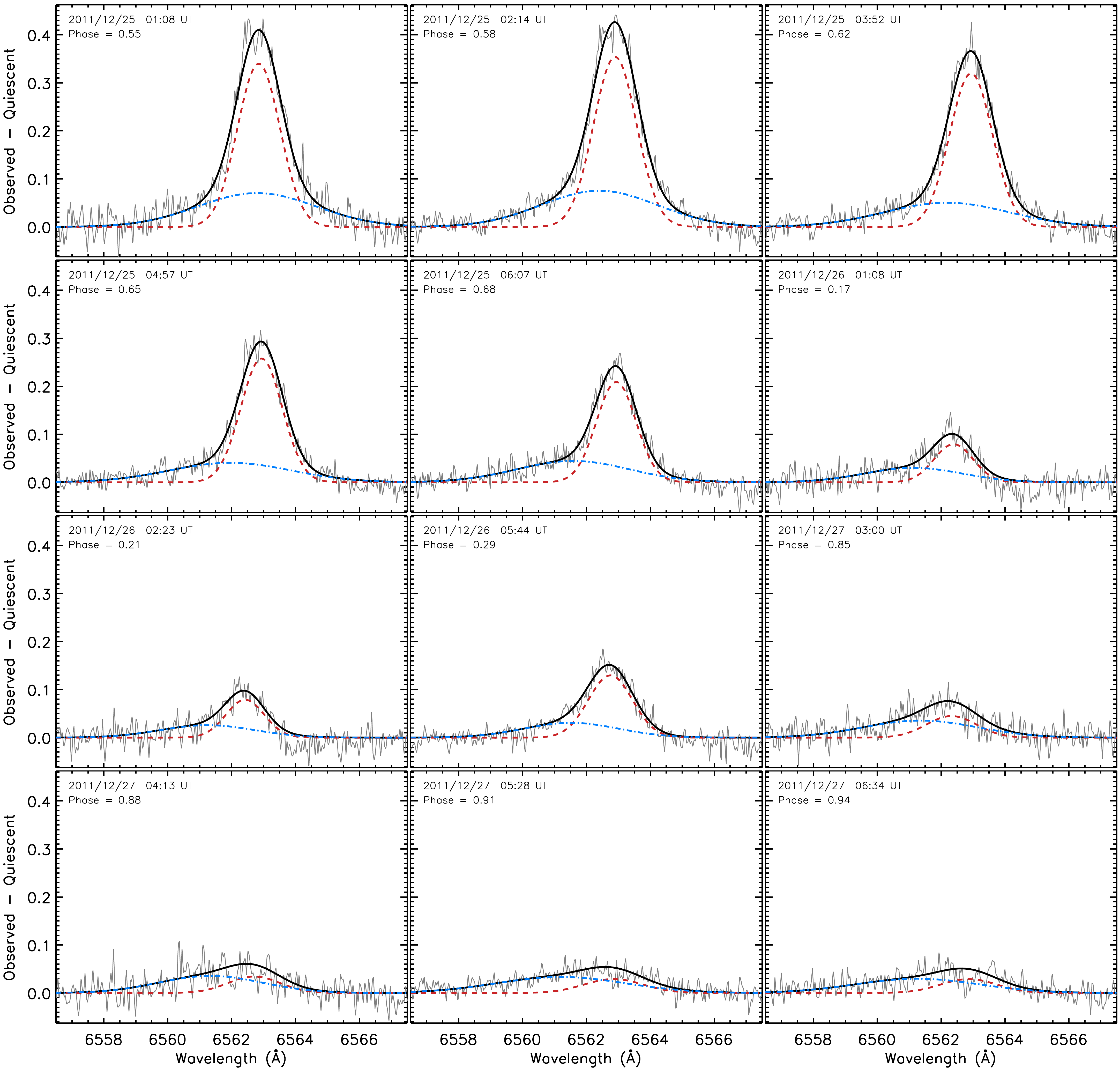}

\caption{Two Gaussian fit (thick solid) of the subtracted H$\alpha$ profiles
(gray solid) of a flare and post-flare spectra. The narrow and broad
Gauassian components are depicted as a red dashed, and as a blue dash-dotted
line, respectively. Time proceeds from top left to bottom right.\label{fig:Ha2G}}
\end{figure*}

As the original purpose of our observations was the study of the photospheric
spot evolution on a time scale of several months, rather than the
analysis of flares, the cadence and integration time with which our
spectra were collected are generally not optimal for a detailed study
of flare evolution. The exception to this was the long duration flare
detected the second day of observations (2011 December 25) and the
associated post-flare activity observed during the two following days.
To study the variations of the line profiles and their asymmetries,
we subtracted a quiescent spectrum defined as the average of the spectra
with $F_{\mathrm{H\alpha}}\leq3.5\times10^{6}\,\mathrm{erg\,cm^{-2}\,s^{-1}}$
($EW_{\mathrm{H\alpha}}\apprle0.8$\,\AA). The reason of taking
an average profile as reference is to reduce the noise of the template,
and the effect that telluric lines and possible rotational modulations
could have on the line asymmetries. The resulting evolution of the
profiles can be seen in Fig. \ref{fig:Ha2G}. In \citetalias{2015A&A...575A..57F}
we evaluated the EW of the net emission profiles after the subtraction
of the spectrum of a non-active star. We refrain here from using the
same spectrum because it introduces some weak asymmetries in the profiles.

\begin{table*}
\caption{Parameters of the subtracted H$\alpha$ lines and their broad and
narrow Gaussian components during the flare and post-flare.\label{tab:Tab_2Gauss}}

\begin{centering}
\begin{tabular}{ccccccS[table-format=+2.0,table-figures-uncertainty=1]ccS[table-format=+2.1,table-figures-uncertainty=1]S[table-figures-uncertainty=1]c}
\hline\hline
\noalign{\vskip0.1cm} &  &  &  &  & \multicolumn{3}{c}{Broad component} &  & \multicolumn{3}{c}{Narrow component}\tabularnewline
\cline{6-8} \cline{10-12} 
\noalign{\vskip0.15cm}
$\mathrm{HJD}-\mathrm{HJD_{0}}$ & Phase & $\mathrm{EW}$ & {\large{}$\nicefrac{\chi_{_{\mathrm{1G}}}^{2}}{\chi_{_{\mathrm{2G}}}^{2}}$} &  & $I\times10^{-2}$ & {$\Delta\lambda$} & FWHM &  & {$I\times10^{-2}$} & {$\Delta\lambda$} & FWHM\tabularnewline
(days) &  & (\AA) &  &  &  & {($\mathrm{km\thinspace s^{-1}}$)} & (\AA) &  &  & {($\mathrm{km\thinspace s^{-1}}$)} & (\AA)\tabularnewline
\hline 
\noalign{\vskip0.1cm}
0.88 & 0.55 & $1.66\pm0.02$ & 1.4 &  & $7.0\pm0.3$ & +0\pm2 & $4.4\pm0.1$ &  & 34.0\pm0.3 & +3\pm1 & $1.57\pm0.02$\tabularnewline
\noalign{\vskip0.1cm}
0.92 & 0.58 & $1.69\pm0.01$ & 1.9 &  & $7.5\pm0.3$ & -18\pm3 & $4.3\pm0.1$ &  & 35.4\pm0.4 & +5\pm1 & $1.58\pm0.02$\tabularnewline
\noalign{\vskip0.1cm}
0.99 & 0.62 & $1.53\pm0.01$ & 1.6 &  & $5.0\pm0.3$ & -28\pm6 & $4.6\pm0.1$ &  & 31.9\pm0.4 & +7\pm1 & $1.58\pm0.01$\tabularnewline
\noalign{\vskip0.1cm}
1.04 & 0.65 & $1.36\pm0.01$ & 1.6 &  & $4.0\pm0.3$ & -38\pm6 & $4.5\pm0.1$ &  & 25.8\pm0.4 & +7\pm1 & $1.52\pm0.03$\tabularnewline
\noalign{\vskip0.1cm}
1.09 & 0.68 & $1.26\pm0.01$ & 1.9 &  & $4.4\pm0.2$ & -55\pm5 & $4.2\pm0.1$ &  & 20.9\pm0.3 & +7\pm1 & $1.46\pm0.02$\tabularnewline
\noalign{\vskip0.1cm}
1.88 & 0.17 & $0.94\pm0.01$ & 1.1 &  & $3.0\pm0.3$ & -76\pm9 & $3.6\pm0.2$ &  & 7.9\pm0.5 & -18\pm3 & $1.44\pm0.09$\tabularnewline
\noalign{\vskip0.1cm}
1.93 & 0.21 & $0.92\pm0.01$ & 1.0 &  & $2.6\pm0.3$ & -71\pm9 & $3.5\pm0.2$ &  & 7.9\pm0.4 & -17\pm2 & $1.42\pm0.07$\tabularnewline
\noalign{\vskip0.1cm}
2.07 & 0.29 & $1.06\pm0.01$ & 1.1 &  & $3.1\pm0.4$ & -58\pm12 & $3.5\pm0.3$ &  & 13.0\pm0.5 & -2\pm1 & $1.63\pm0.04$\tabularnewline
\noalign{\vskip0.1cm}
2.96 & 0.85 & $1.00\pm0.01$ & 1.0 &  & $3.5\pm0.5$ & -66\pm10 & $4.2\pm0.4$ &  & 4.5\pm0.5 & -21\pm9 & $1.85\pm0.12$\tabularnewline
\noalign{\vskip0.1cm}
3.01 & 0.88 & $0.94\pm0.02$ & 0.9 &  & $3.6\pm0.4$ & -67\pm10 & $3.8\pm0.3$ &  & 3.4\pm0.5 & -4\pm11 & $1.91\pm0.18$\tabularnewline
\noalign{\vskip0.1cm}
3.06 & 0.91 & $0.99\pm0.01$ & 1.0 &  & $3.3\pm0.3$ & -73\pm12 & $4.8\pm0.4$ &  & 2.9\pm0.4 & +3\pm11 & $2.17\pm0.19$\tabularnewline
\noalign{\vskip0.1cm}
3.10 & 0.94 & $0.96\pm0.01$ & 1.0 &  & $3.0\pm0.3$ & -70\pm13 & $4.6\pm0.3$ &  & 2.9\pm0.4 & +3\pm13 & $2.08\pm0.22$\tabularnewline
\hline 
\end{tabular}
\par\end{centering}
\tablefoot{Times are relative to $\mathrm{HJD_{0}}=2455919.67$.
The equivalent widths are from \citetalias{2015A&A...575A..57F} and
were evaluated after the subtraction of a non-active template. The
line shifts $\Delta\lambda$ are relative to $\lambda_{\mathrm{H\alpha}}=6562.79$\,\AA.}
\end{table*}

The subtracted profiles show enhanced emission in the wings of the
lines. This is a well known effect of flares on Balmer lines, and
has already been observed by several authors (see, e.g., \citealt{1987LNP...291..173D};
\citealt{1988A&A...193..229D}; \citealt{1988MNRAS.235..573P}; \citealt{1992AJ....104.1161E}),
also in LQ~Hya \citep{1999MNRAS.305...45M}. Single-Gaussian fits
of these profiles give a poor description of the lines, and a second,
broader, and usually shifted Gaussian is necessary for an acceptable
fit. To evaluate the parameters of the Gaussians and their uncertainties,
we fit each subtracted profile one thousand times with a set of randomly
generated initial conditions. The solution for each line is then the
weighted mean of all these evaluations, while their uncertainties
are the weighted standard deviation. As weighting factor we used the
chi-square goodness of the fit. This process is not really necessary
when the emission is high because the solution is then well constrained.
Nevertheless, when the emission is low the problem becomes ill-posed
and many solutions are possible within the uncertainties. The resultant
two-Gaussian fit of the profiles, with a broad and a narrow component,
is shown in Fig. \ref{fig:Ha2G}. The numerical values are available
in Table~\ref{tab:Tab_2Gauss}, where we also indicate the improvement
in using two Gaussians rather than one as the ratio of the chi-square
goodness of the fits $\nicefrac{\chi_{_{\mathrm{1G}}}^{2}}{\chi_{_{\mathrm{2G}}}^{2}}$.

The spectroscopic observations of the flare consist of five spectra
collected with a cadence somewhat higher than one hour. All of them
were taken near rotational phase 0.6, with the active hemisphere of
the star in its best view. The highest H$\alpha$ emission is detected
in the second spectrum of the night. The H$\alpha$ EW of the first
spectrum is just 0.03\,\AA\ below the maximum, and could correspond
to the very end of the impulsive phase. The remaining three spectra
show the slow decay of the flare. When the observations finished that
night, the EW was 1.26\,\AA, still 26\% above the basal emission.
The broad component of the subtracted H$\alpha$ profile also shows
clear variations. The first spectrum has its both Gaussians practically
aligned at $\lambda_{\mathrm{H\alpha}}=6562.79$\,\AA, but afterwards
the broad component moves toward shorter wavelengths. The last spectrum
of the flare shows the broad Gaussian blue-shifted by $\approx1.2$\,\AA\
($55\mathrm{\thinspace km\thinspace s^{-1}}$) relative to $\lambda_{\mathrm{H\alpha}}$. 

Although the flare was not observed in its full extension, a lower
limit of the energy released in the H$\alpha$ and H$\beta$ lines
can be estimated by converting the radiative fluxes from \citetalias{2015A&A...575A..57F}
into luminosities and integrating them over time. To isolate the contribution
of the flare from the basal emission we assumed $F_{\mathrm{H\alpha}}=4.3\times10^{6}\,\mathrm{erg\,cm^{-2}\,s^{-1}}$
and $F_{\mathrm{H\beta}}=1.6\times10^{6}\,\mathrm{erg\,cm^{-2}\,s^{-1}}$
as the amount of flux received from the undisturbed chromosphere.
This leads to the energies $E_{\mathrm{H\alpha}}>5.2\times10^{32}\,\mathrm{erg}$
and $E_{\mathrm{H\beta}}>3.1\times10^{32}\,\mathrm{erg}$ if a radius
$R_{\mathrm{star}}=0.97\thinspace R_{\textrm{\ensuremath{\astrosun}}}$
is adopted \citep{2004A&A...417.1047K}. In comparison, \citet{1999MNRAS.305...45M}
measured $E_{\mathrm{H\alpha}}>2.5\times10^{33}\,\mathrm{erg}$ and
$E_{\mathrm{H\beta}}>2.1\times10^{33}\,\mathrm{erg}$ for a different
LQ Hya flare.

One day after the flare the star was observed near rotational phase
0.2. This corresponds mostly to the most quiet hemisphere of the star,
but with the active hemisphere already visible near the limb. At that
time the EW of the H$\alpha$ line was near 0.9\,\AA, which is within
the usual range of values for the basal emission at that phase. But
even though the H$\alpha$ emission is comparably low, the line shape
had a blue wing with excess emission. The broad Gaussian component
of these profiles is weaker than in the spectra taken the day before
and also slightly more blue-shifted than in the last flare spectrum.
Considering that we did not observe this kind of behavior in any other
spectra from the same phase, we speculate that it could be related
to the flare observed the day before. The last spectrum of the night
was taken at phase 0.3 and shows a small increment in the emission.
It is likely because of the increment of the projected area of the
active regions appearing from the limb, but could have been also due
to a minor flare.

\begin{figure}
\includegraphics[width=1\columnwidth]{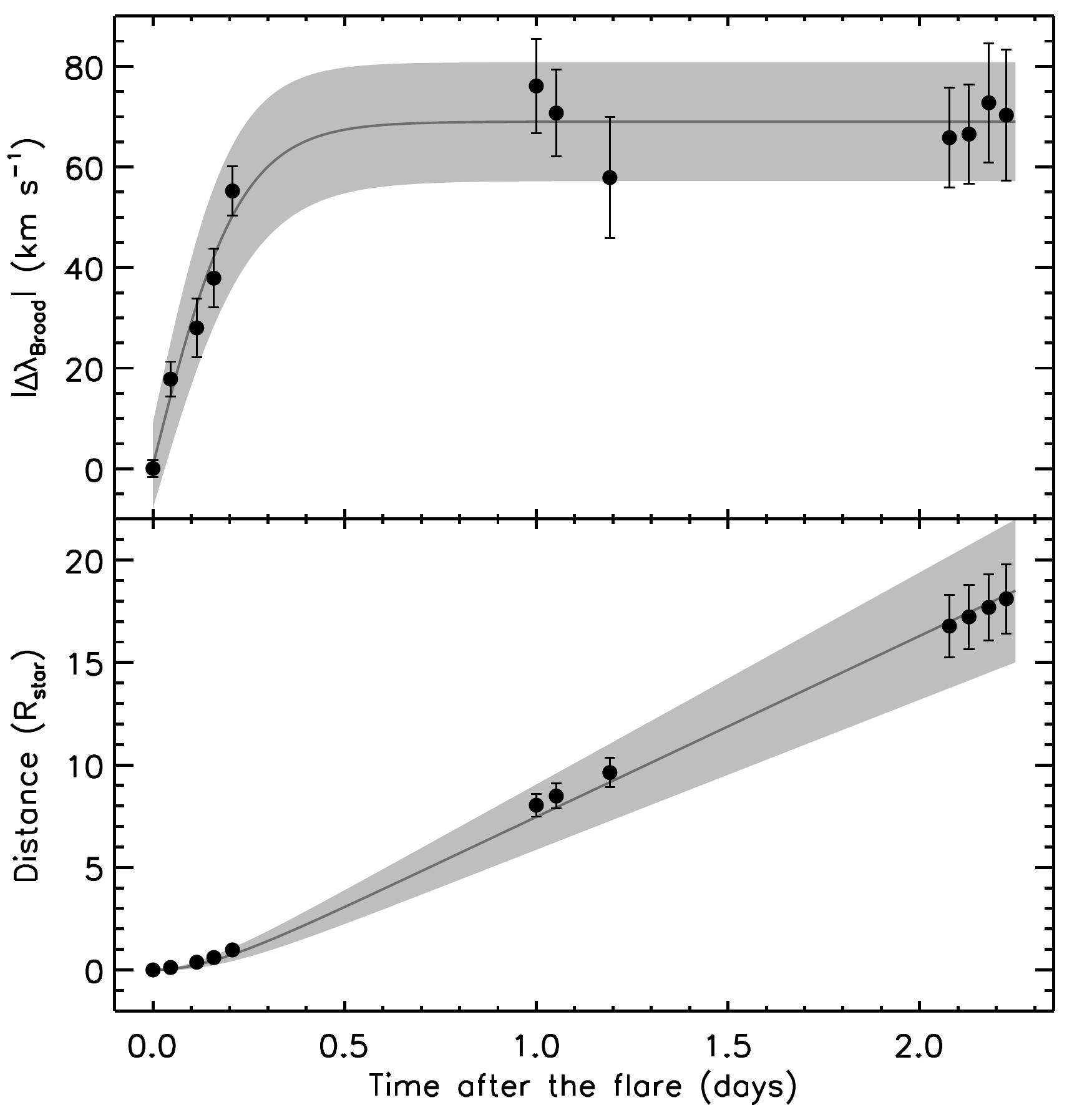}

\caption{\emph{Top:} absolute shift of the broad Gaussian component of the
H$\alpha$ line from Fig.~\ref{fig:Ha2G}, in velocity scale. The
data have been fit to Eq.~(\ref{eq:velo}) (solid line). The gray
area indicates the 95\% confidence intervals of the fit. \emph{Bottom:}
distance traveled by the source of the broad component under the assumption
that it corresponds to ejected material. Data points are from the
numerical integration of the velocity measurements. The solid line
is the integral of the function used to fit the velocities. \label{fig:DeltLamb}}
\end{figure}

Two days after the strong flare the blue wing of the H$\alpha$ line
still presents excess emission. Four spectra taken near phase 0.9
are available from that night. The emission from the core was then
clearly lower that the day before, but the emission of both wings
was slightly enhanced. This could be a physical process or maybe a
somewhat misplaced continuum. A single-Gaussian fit of the subtracted
profiles provides here an acceptable result. Nevertheless, to look
for a continuity in the behavior of the broad component, we forced
our code to use two Gaussians, one of them as a narrow component with
center near $\lambda_{\mathrm{H_{\alpha}}}$. By doing so, we find
broad components that are wider than the day before but placed at
the same position. The time evolution of the line shifts in velocity
scale is shown in the top panel of Fig.~\ref{fig:DeltLamb}.

The asymmetries of the Balmer lines are usually interpreted as a consequence
of plasma turbulence or directed mass motions. \citet{1990A&A...238..249H}
found a large enhancement in the far blue wings of Balmer lines during
the impulsive phase of a flare on AD~Leo, that they attribute to
a mass ejection. \citet{1993A&A...274..245H} found in the same flare
red-shifted Balmer line cores that were interpreted as a chromospheric
downward condensation. Blue-shifted asymmetries in the Balmer lines
of AT~Mic were interpreted by \citet{1994A&A...285..489G} as a result
of a high-velocity chromospheric evaporation during a flare. \citet{1999MNRAS.305...45M}
performed a multi-line analysis of an unusually strong flare on LQ~Hya
and detected asymmetric profiles that were blue-shifted during the
impulsive phase and red-shifted during the gradual decay. They attribute
the broad components and the asymmetries to plasma turbulence or to
upward and downward mass motions.

\begin{table*}
\centering
\caption{Time interval covered by each Doppler map. \label{tab:Time-intervals} }

\begin{tabular}{ccccccc}
\hline\hline
\noalign{\vskip0.1cm} & \multicolumn{2}{c}{Time range} &  &  &  & \tabularnewline
\cline{2-3} 
\noalign{\vskip0.15cm}
Group & $\mathrm{HJD}-\mathrm{HJD_{0}}$ & Date & Days & Spectra & Spectra per map & Photom. points\tabularnewline
\hline 
\noalign{\vskip0.1cm}
1 & \hphantom{0}0 - 21 & 2011/12/24 - 2012/01/15 & 22 & 56 & 29 & 34\tabularnewline
\noalign{\vskip0.1cm}
2 & 22 - 53 & 2012/01/15 - 2012/02/16 & 32 & 55 & 47 & 57\tabularnewline
\noalign{\vskip0.1cm}
3 & 54 - 80 & 2012/02/16 - 2012/03/14 & 27 & 41 & 30 & 626\tabularnewline
\noalign{\vskip0.1cm}
4 & \hphantom{0}81 - 102 & 2012/03/14 - 2012/04/05 & 22 & 22 & 19 & 179\tabularnewline
\noalign{\vskip0.1cm}
5 & 103 - 129 (+11) & 2012/04/05 - 2012/04/30 (+11\,d) & 27 (+11) & 25 & 13 & 248\tabularnewline
\hline 
\end{tabular}

\tablefoot{To facilitate the comparison, we divided the photometry
from Fig.~\ref{fig:Photometry} in the same five groups. The last
group covers for the photometry eleven additional days for which we
do not have spectroscopic data. $\mathrm{HJD_{0}}$ is the same as
in Tab. \ref{tab:Tab_2Gauss}.}
\end{table*}

Although the measurements are too sparse in time to conclude beyond
all doubt that the excess emission in the blue wing of the H$\alpha$
line is the consequence of a single event, we speculate that the source
could be some kind of material ejected during the flare. The excess
emission was detected in the H$\alpha$ line but not in H$\beta$,
which could indicate that the emission came from optically thin material
seen off-limb. Additionally, it was observed at different rotational
phases, for as long as two days after the flare, and always blue-shifted.
This lack of rotational modulation could suggest that the source of
emission is not attached to the stellar surface and that it is not
corotating with it. Furthermore, the velocity profile calculated from
the line shifts (top panel of Fig.~\ref{fig:DeltLamb}) is remarkably
similar to that of many solar coronal mass ejections (see, e.g., \citealt{2003ApJ...588L..53G};
\citealt{2011ApJ...738..191B}).

Under the hypothesis of the mass ejection, we estimated the distance
that the material could have traveled. We first fit the velocities
to a function of the form 
\begin{equation}
v(t)=\frac{a}{\mathrm{e}^{-b\cdot t}+1}-c,\label{eq:velo}
\end{equation}
 where \emph{t} is time, and $a$, $b$ and $c$ are the fitting parameters.
The function was chosen because it reproduces well the behavior of
the data, with a fast and short acceleration phase, and a long asymptotic
phase of constant velocity. The gray areas of Fig. \ref{fig:DeltLamb}
are the 95\% confidence intervals of the fit. Once the fitting function
was evaluated, we determined the distances by just integrating it
(lower panel of Fig. \ref{fig:DeltLamb}). According to this, the
distance traveled would be of $\approx18\thinspace R_{\mathrm{star}}$,
where we took $R_{\mathrm{star}}=0.97\thinspace R_{\textrm{\ensuremath{\astrosun}}}$
from \citet{2004A&A...417.1047K}. We note that because of the use
of projected velocities, this distance is only a lower limit.

\begin{figure*}
\includegraphics[width=1\textwidth]{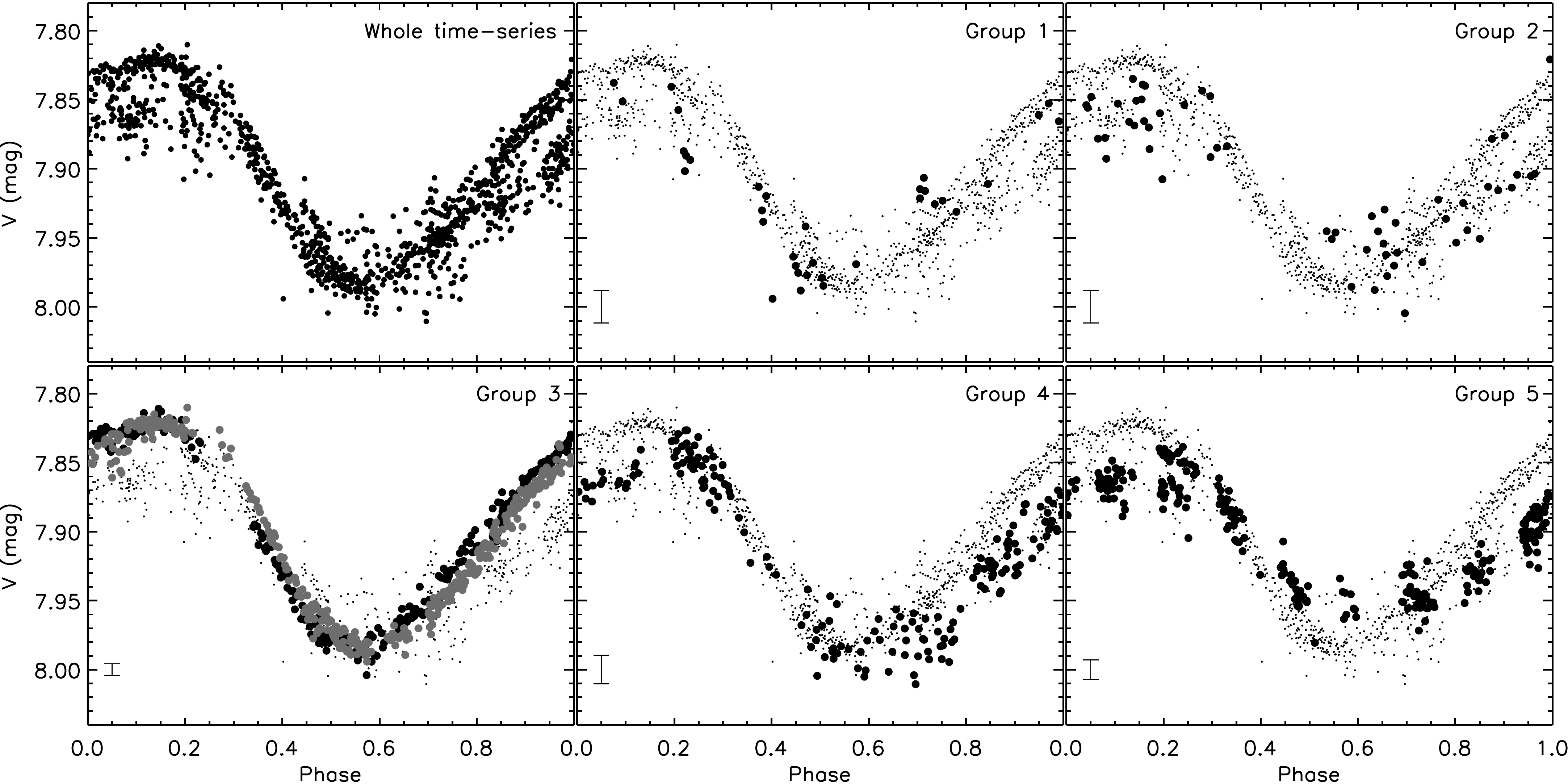}

\caption{Rotational modulation of the \emph{V}-band photometry at different
time intervals. \emph{Top left:} for the whole time-series. \emph{Top
middle:} for days 0 to 21. \emph{Top right:} for days 22 to 53. \emph{Bottom
left:} for days 54 to 66 (black circles) and for days 67 to 80 (gray
circles). \emph{Bottom middle:} for days 81 to 102. \emph{Bottom right:}
for days 103 to 140. Average error bars for each group are given in
the lower left corner of each panel. Small dots correspond to the
measurements of the whole time-series and are included as a reference
to facilitate the comparison. Times in calendar dates are available
in Table~\ref{tab:Time-intervals}.\label{fig:Photometry}}
\end{figure*}

We conclude this section with a brief summary about the H$\alpha$
line asymmetries in the rest of the spectra. The two Gaussian fits
of the H$\alpha$ lines with $EW\lesssim1.0$\,\AA\ do not show
any improvement over the use of a single Gaussian. On the other hand,
the use of two Gaussians improves the fit of all H$\alpha$ lines
with $EW\apprge1.1$\,\AA\ by at least 10\% in terms of the chi-square
goodness of the fit. When the EW is in the range 1.0-1.1\,\AA\ roughly
half of the fits show some improvement if two Gaussians are used.
We note, however, that these values depend on the selection of the
``quiescent'' spectrum. Approximately 40 spectra show H$\alpha$
lines with a shifted broad component, most of them red-shifted. 

\citet{1998ASPC..154.1516M} analyzed the H$\alpha$ line in chromospherically
active binary systems and in weak-lined T~Tauri stars. They detected
in the most active stars of the sample broad wings that in some cases
were also asymmetric. By similarity with the behavior of transition
region lines, they interpreted this excess emission as arising from
microflare activity. We have found a comparable behavior in the lines
from the most active hemisphere of LQ~Hya. This suggest that at least
part of the basal emission from that side of the star could be produced
by low intensity flares occurring on a regular basis, supporting our
results from \citetalias{2015A&A...575A..57F}. A few examples of
these lines are shown in the appendix in Fig.~\ref{fig:Example-of-three}.

\section{Evolution of the photospheric spots \label{sec:Evo_fotosfera}}

\subsection{Photometric light curves \label{sec:Photometry}}

To facilitate the comparison between the photometric light curve and
the Doppler maps of Sect.~\ref{sec:Doppler-imaging}, we have divided
them into the same five time intervals (Tab.~\ref{tab:Time-intervals}).
The photometry of the last group covers eleven additional days for
which we do not have any spectral data. Data collected during the
first 58 days of observation (first and second groups) are affected
by larger error bars. Nevertheless, we include the best measurements
for completeness and because they could still provide some guidelines
about the evolution. More than a half of the photometric data were
taken during the time span covered by the third group. This high number
of measurements together with their significantly smaller error bars,
make possible to see the evolution at a shorter time scale. For these
reasons we have divided this group into two subgroups.

The top left panel of Fig.~\ref{fig:Photometry} shows the phase-plot
of the whole time-series. As reported in \citetalias{2015A&A...575A..57F},
the photometry shows clear rotational modulation, indicating a higher
concentration of dark spots on one side of the star. Although this
configuration is relatively stable in time, it is still possible to
see some evolution. In the third group (bottom left panel of Fig.~\ref{fig:Photometry})
we observe that the time of minimum brightness moves toward higher
phases. As for the chromospheric emission, this can be easily explained
if the rotation period of the spots were slightly longer than the
$P=1.60066$\,d period used to phase the data. Nevertheless, subsequent
data show that this can not be the only reason. We see that while
the hemisphere of minimum brightness moves from phase \textasciitilde{}0.5
to phase \textasciitilde{}0.6, the peak-to-peak variation of the light
curve decreases from 0\fm16 in the third group to 0\fm09 at the
end of the observations. Contrary to what we observe for the basal
H$\alpha$ emission, the $V$ brightness of the less active side of
the star takes always values that are clearly different from those
of the opposite hemisphere. We also note that the changes in the light
curve take place while the mean brightness of the star remains roughly
constant at a value of $V\approx7.90$\,mag. This indicates that
the main reason of variability could be the redistribution of spots
on the stellar surface, rather than the emerging of new or decay of
old active regions. Finally, the unspotted magnitude of LQ~Hya is
thought to be near $V=7.76$\,mag (see, e.g., \citealt{1993A&A...268..671S}),
which is clearly brighter than our maximum value of $V=7.82$\,mag.
This suggests that not even the most quiet hemisphere of LQ~Hya was
completely spotless during our observations.

\subsection{Doppler imaging \label{sec:Doppler-imaging}}

\subsubsection{Line inversion}

The chromospheric H$\alpha$ emission and the photometric light curve
clearly indicate that the surface activity of LQ Hya was changing
over time. As they alone can not show the ultimate cause of the changes,
we have spatially resolved the surface of the star by means of Doppler
imaging. We inverted the line profiles with the algorithm developed
by \citet{1987Geop...52..289C} and \citet{1990Geop...55.1613D}.
The local line profiles were computed with the program SPECTRUM \citep{1994AJ....107..742G}
for a grid of $60\times30$ surface elements. We used the model atmospheres
from \citet{1993KurCD..13.....K}, the stellar parameters from \citet{2004A&A...417.1047K},
and the chemical abundances from \citet{1998A&A...336..972R} (Table
\ref{tab:LQ_Hya_Parameters}). The result of the inversions are shown
in Fig.~\ref{fig:Maps}.

\begin{table}
\caption{Parameters of LQ Hya adopted for the inversion of the line profiles.\textbf{
}\label{tab:LQ_Hya_Parameters}}

\begin{centering}
\begin{tabular}{lll}
\hline\hline
\noalign{\vskip0.1cm} Parameter &  & Value\tabularnewline
\hline 
\noalign{\vskip0.06cm}
{\small{}Spectral type} &  & {\small{}K2\,V}\tabularnewline
\noalign{\vskip0.06cm}
$\log g$ &  & {\small{}4.0}\tabularnewline
\noalign{\vskip0.06cm}
$v\sin i$ &  & {\small{}$28.0\thinspace\mathrm{km\thinspace s^{-1}}$}\tabularnewline
\noalign{\vskip0.06cm}
Inclination,\emph{ i} &  & 65\degree\tabularnewline
\noalign{\vskip0.06cm}
Rotation period, $P_{\mathrm{rot}}$ &  & 1.60066\,d\tabularnewline
\noalign{\vskip0.06cm}
Radius, \emph{R} &  & {\small{}0.97}\tabularnewline
\noalign{\vskip0.06cm}
Microturbulence &  & {\small{}$0.5\thinspace\mathrm{km\thinspace s^{-1}}$}\tabularnewline
\noalign{\vskip0.06cm}
Macroturbulence &  & {\small{}$1.5\thinspace\mathrm{km\thinspace s^{-1}}$}\tabularnewline
\noalign{\vskip0.06cm}
Fe abundance &  & {\small{}solar + 0.1\,dex}\tabularnewline
\noalign{\vskip0.06cm}
Ca abundance &  & {\small{}solar + 0.1\,dex}\tabularnewline
\noalign{\vskip0.06cm}
Ni abundance &  & {\small{}solar + 0.3\,dex}\tabularnewline
\noalign{\vskip0.06cm}
Abund. other elements &  & {\small{}solar}\tabularnewline
\hline 
\end{tabular}
\par\end{centering}
\tablefoot{The physical parameters are from \citet{2004A&A...417.1047K}
and the chemical abundances from \citet{1998A&A...336..972R}.}
\end{table}

\begin{table}
\caption{Spectral lines used for the inversion, together with the adopted $\log gf$
and energy of the lower state. \label{tab:Inversion-lines}}

\centering{}%
\begin{tabular}{ccc>{\centering}p{0.3cm}S[table-format=1.3]>{\centering}p{0.3cm}S[table-format=5.0]}
\hline\hline
\noalign{\vskip0.1cm} {\small{}Element-Ion} &  & {\small{}$\lambda_{\mathrm{cent}}$} &  & {\small{}{$\log gf$}} &  & {\small{}{$E_{\mathrm{low}}$}}\tabularnewline
 &  & {\small{}(\AA)} &  &  &  & {\small{}{($\mathrm{cm^{-1}}$)}}\tabularnewline
\hline 
{\small{}\ion{Fe}{I}} &  & {\small{}5429.696} &  & {\small{}-1.879} &  & {\small{}7728}\tabularnewline
{\small{}\ion{Fe}{I}} &  & {\small{}5501.464} &  & {\small{}-2.957} &  & {\small{}7728}\tabularnewline
{\small{}\ion{Fe}{I}} &  & {\small{}5569.618} &  & {\small{}-0.633} &  & {\small{}27560}\tabularnewline
{\small{}\ion{Fe}{I}} &  & {\small{}5572.841} &  & {\small{}-0.311} &  & {\small{}27395}\tabularnewline
{\small{}\ion{Fe}{I}} &  & {\small{}5576.090} &  & {\small{}-1.01} &  & {\small{}27666}\tabularnewline
{\small{}\ion{Ca}{I}} &  & {\small{}5581.965} &  & {\small{}-0.510} &  & {\small{}20349}\tabularnewline
{\small{}\ion{Ca}{I}} &  & {\small{}5594.462} &  & {\small{}0.15} &  & {\small{}20349}\tabularnewline
{\small{}\ion{Ca}{I}} &  & {\small{}5598.480} &  & {\small{}0.034} &  & {\small{}20335}\tabularnewline
{\small{}\ion{Ca}{I}} &  & {\small{}5857.451} &  & {\small{}0.257} &  & {\small{}23652}\tabularnewline
{\small{}\ion{Fe}{I}} &  & {\small{}5914.194} &  & {\small{}-0.059} &  & {\small{}37163}\tabularnewline
{\small{}\ion{Fe}{I}} &  & {\small{}6024.049} &  & {\small{}0.091} &  & {\small{}36686}\tabularnewline
{\small{}\ion{Fe}{I}} &  & {\small{}6191.558} &  & {\small{}-1.377} &  & {\small{}19621}\tabularnewline
{\small{}\ion{Fe}{I}} &  & {\small{}6400.009} &  & {\small{}-0.290} &  & {\small{}29056}\tabularnewline
{\small{}\ion{Fe}{I}} &  & {\small{}6411.658} &  & {\small{}-0.718} &  & {\small{}29469}\tabularnewline
{\small{}\ion{Fe}{I}} &  & {\small{}6430.856} &  & {\small{}-2.006} &  & {\small{}17550}\tabularnewline
{\small{}\ion{Ca}{I}} &  & {\small{}6439.081} &  & {\small{}0.470} &  & {\small{}20371}\tabularnewline
{\small{}\ion{Ca}{I}} &  & {\small{}6462.570} &  & {\small{}0.30} &  & {\small{}20349}\tabularnewline
{\small{}\ion{Ca}{I}} &  & {\small{}6717.688} &  & {\small{}-0.24} &  & {\small{}21850}\tabularnewline
{\small{}\ion{Ca}{I}} &  & {\small{}7148.150} &  & {\small{}0.208} &  & {\small{}21850}\tabularnewline
\hline 
\end{tabular}
\end{table}

Particularly problematic for the inversion was the rather high level
of noise in the spectra. The spots of LQ~Hya have only a relatively
weak effect on the line profiles (see Fig.~\ref{fig:Line-fits} in
the appendix), and an average S/N of 74 is certainly far lower than
ideal for Doppler imaging. To minimize this problem, we inverted simultaneously
19 line profiles (Table~\ref{tab:Inversion-lines}), whose level
of noise was previously reduced with the PCA denoising procedure described
by \citet{2008A&A...486..637M}. Neighbor lines affecting those profiles
were also taken into account for the inversion. In addition to the
noise, magnetic activity phenomena other than photospheric spots also
seem to be affecting the shape of the spectral lines. We noted that
several line profiles are different to others taken at a similar phase
and time. This is particularly evident at the beginning of the observations,
when the flare activity was more intense. Typical effects are the
line filling produced during flares, and line asymmetries that could
be attributed to the transfer of momentum to the photosphere by down-flows
similar to those observed in the Sun (see, e.g., \citealt{1998Natur.393..317K};
\citealt{2008MNRAS.387L..69V}). Nevertheless, in most cases it is
not possible to find the ultimate cause of the discrepancies, as noise
itself can have a similar effect. Line profiles affected by these
problems were not used in the inversion.

An indirect way to reduce the effect of noise and activity is to use
a high amount of spectra per map. To avoid inconsistencies introduced
by the natural evolution of active regions, this has to be done using
data that cover sufficiently short time intervals. For the computation
of the maps we divided the spectra in the five groups indicated in
Table~\ref{tab:Time-intervals}. With this distribution of spectra
we try to have always a good phase coverage, while keeping all groups
with a similar length. Nonetheless, this was not always possible because
of the nonuniform sampling of our data, and the presence of gaps.
We note that the quality of the last map is comparably lower than
for the others as a consequence of the poor phase coverage near phase
0.5. Furthermore, the few spectra available from that phase are either
affected by a flare or by a low signal to noise ratio.

As an example, we show in the appendix in Fig.~\ref{fig:Map_G1SpcCME}
a map computed in the same way as the first map of the time series
(top panel of Fig.~\ref{fig:Maps}) but including some spectra that
were rejected for the final solution. These spectra are those in which
the H$\alpha$ line seems to be affected by a mass ejection (see Sect.~\ref{subsec:-ME?}),
but that are undisturbed by the flare that could have produced it
(i.e., the last seven spectra of Table~\ref{tab:Tab_2Gauss}). In
this version of the map the near equatorial spot is no longer present,
 the temperature of the bridge between the two high latitude spots
decreases $\approx120$\,K to values near those from their cores,
and a new spot appears near phase 0.85. Nonetheless, our data suggest
that this new spot is not a real cool spot. It appears on the least
active side of the star at a time when the chromospheric emission
had its lowest values, it does not seem to have any effect on the
photometry, and it is not observed in spectra that were taken just
three days after and three days before than the rejected spectra nor
in any other map. It can also be noted that after removing the affected
spectra a 0.2 phases wide gap appears near phase 0.85. This gap is
nonetheless not big enough to explain the absence of the spot. Because
of its high latitude, if the spot were real it should have been detected
by means of the spectra taken within a window of $\pm0.25$ phases.
Although the distortions in the spectral lines that produce the discrepancies
between the two maps are maybe not a consequence of the ejected material
itself, this result suggests that flare activity could affect the
results of line inversions even when the flare emission has already
vanished. We note that the source of these differences could be also
affecting the final version, although in a much slighter degree. It
shows a very weak spot at phase 0.85 and the near equatorial spot
is fainter that in the other maps. 

\subsubsection{Doppler maps}

\begin{figure}[H]
\includegraphics[width=1\columnwidth]{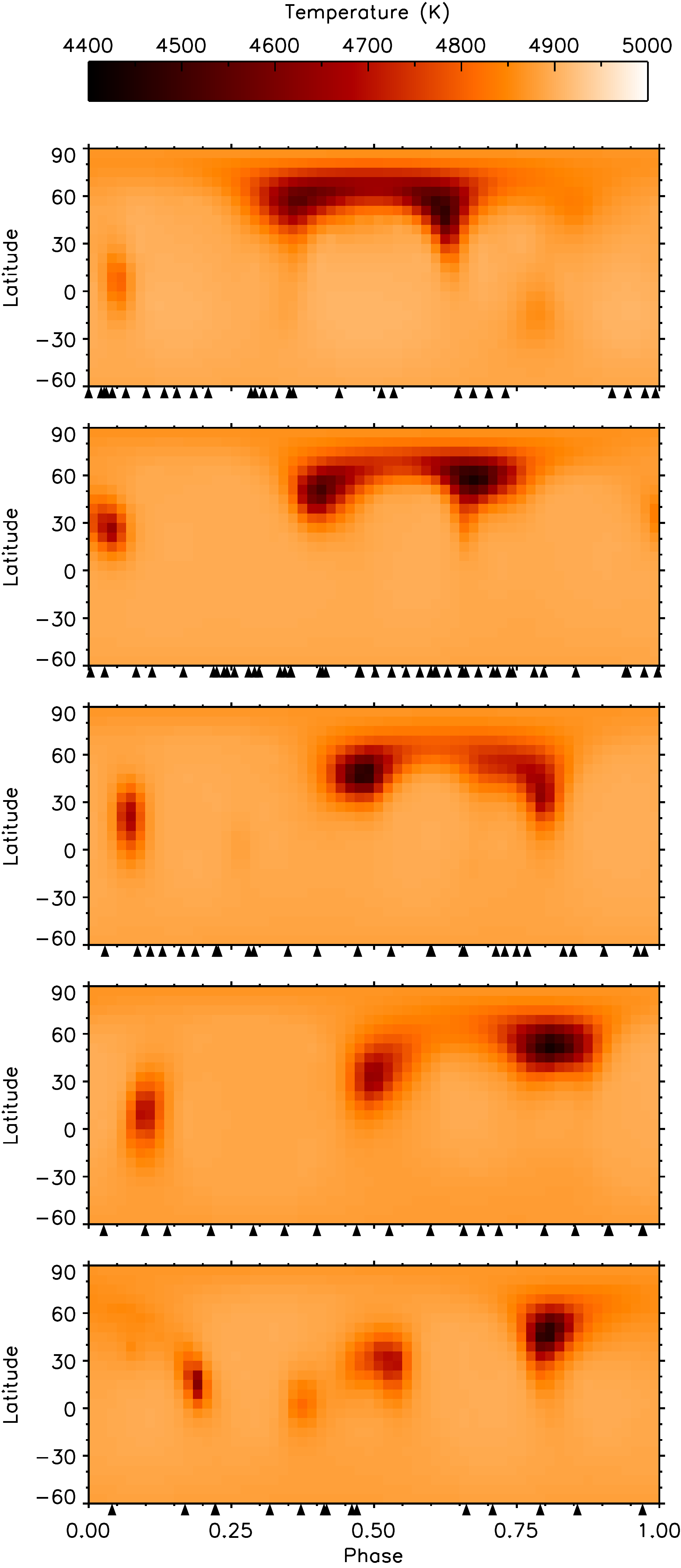}

\caption{Time series of Doppler maps for the period December 2011 - April 2012.
Chronological order from top to bottom. Time intervals covered by
each map are given in Table~\ref{tab:Time-intervals}. Arrows at
the bottom of each panel indicate the phase coverage. Spots are dubbed
S1, S2, and S3 from left to right in the top panel. \label{fig:Maps}}
\end{figure}

Figure~\ref{fig:Maps} reveals that three spots dominated the surface
activity of LQ~Hya during the period of time covered by our observations.
They are a low-latitude spot (S1), and two spots at higher latitudes
(in Fig.~\ref{fig:Maps}, from left to right, S2 and S3). This spot
configuration remained stable and indicates that active regions on
LQ~Hya can easily live for periods of time longer than four months,
with little or no signs of evolution other than changes in their position.
Although some maps also show fluctuation of the spot's temperature
by up to $\approx200$\,K, this could be attributed to the limited
S/N of the data and to flare-related events. For example, the higher
temperature of S3 in the third map seems to be a consequence of inconsistencies
in some line profiles that also lead to a morphological distortion,
and the increment in the temperature of S2 in the fifth map is likely
produced by the poor coverage near phase 0.5. The difference in temperature
between the quiet photosphere and the spots does not exceed here 500\,K,
in good agreement with the results of \citet{1993A&A...268..671S}
and \citet{1998A&A...336..972R}. In the maps the temperature of the
quiet photosphere is near 4900\,K. This is $\approx100$\,K lower
that the maximum hemispheric temperature that we found in \citetalias{2015A&A...575A..57F},
which is not a dramatic difference if we consider that the inversion
has not been optimized for the accurate determination of stellar parameters
other than the position of the spots.

In our maps, all spots show a consistent migration toward higher phases.
Furthermore, we note that this migration is faster at high latitudes
as compared to lower latitudes, as one would expect from a star with
a solar-like differential rotation. From the first map to the fourth
(positions in the fifth map are more uncertain), the low-latitude
spot S1 moves 0.05 phases, S2 moves 0.15 phases, and S3 moves 0.20
phases. The fact that the spots cover a high range of latitudes, and
that all of them move toward higher phases, indicates that the $P=1.60066$\,d
period used to phase the maps could be slightly underestimated. 

Curiously, the latitude of S2 seems to decrease from one map to the
next. From the position of the spot center, and the mid time of each
map, we derive a velocity of $\approx0.26\mathrm{\thinspace deg\thinspace d^{-1}}$.
If we consider the expected radius of a K2 dwarf $R=0.75\thinspace R_{\textrm{\ensuremath{\astrosun}}}$
(e.g., \citealt{2008oasp.book.....G}) and the value $R=0.97\thinspace R_{\textrm{\ensuremath{\astrosun}}}$
found by \citet{2004A&A...417.1047K} for LQ~Hya, this leads to a
meridional speed of $27\mathrm{\thinspace m\thinspace s^{-1}}$ and
$35\thinspace\mathrm{m\thinspace s^{-1}}$, respectively. In comparison,
the meridional motion of sunspot groups rarely exceeds $0.1\mathrm{\thinspace deg\thinspace d^{-1}}$
(see, e.g., \citealt{2001SoPh..198...57W}). Nevertheless, we note
that the determination of latitudes from Doppler imaging is more uncertain
that the determination of rotational phases. Some fluctuation are
therefore normal. This is observed for the spots S1 and S3, as they
irregularly change their latitude by up to 15\,deg. The behavior
of S2 is, on the other hand, somewhat different, as its latitude decreases
monotonically a total of $\approx30$\,deg. A more detailed analysis
of the migration of the spots, including the differential rotation,
is deferred to a future and dedicated paper.

The comparison between the Doppler maps and the photometric light
curve reveals that the phase of minimum brightness corresponds to
the midpoint between the spots S2 and S3. We observe that the motion
of the minimum follows the migration of the two spots toward higher
phases, and that the amplitude of the light curve decreases as the
distance between the spots increases. The darkest side of the star
becomes brighter as S2 and S3 separate from each other, while the
hemisphere centered at phase 0.1 becomes darker as S3 appears from
the limb. S1 seems to play only a minor role in the modulation of
the light curve, surely because of its somewhat higher temperature
and smaller surface. Nevertheless, it explains why we did not observe
the unspotted magnitude of $V\approx7.76$\,mag, as at least one
of the three spots is always on view. This interpretation based purely
on the motion of the spots also explains why the average stellar magnitude
remains constant.

\section{Putting things together: connection between spots, plages, and flares
\label{sec:Putting-things-together}}

\begin{figure*}[!t]
\includegraphics[width=1\textwidth]{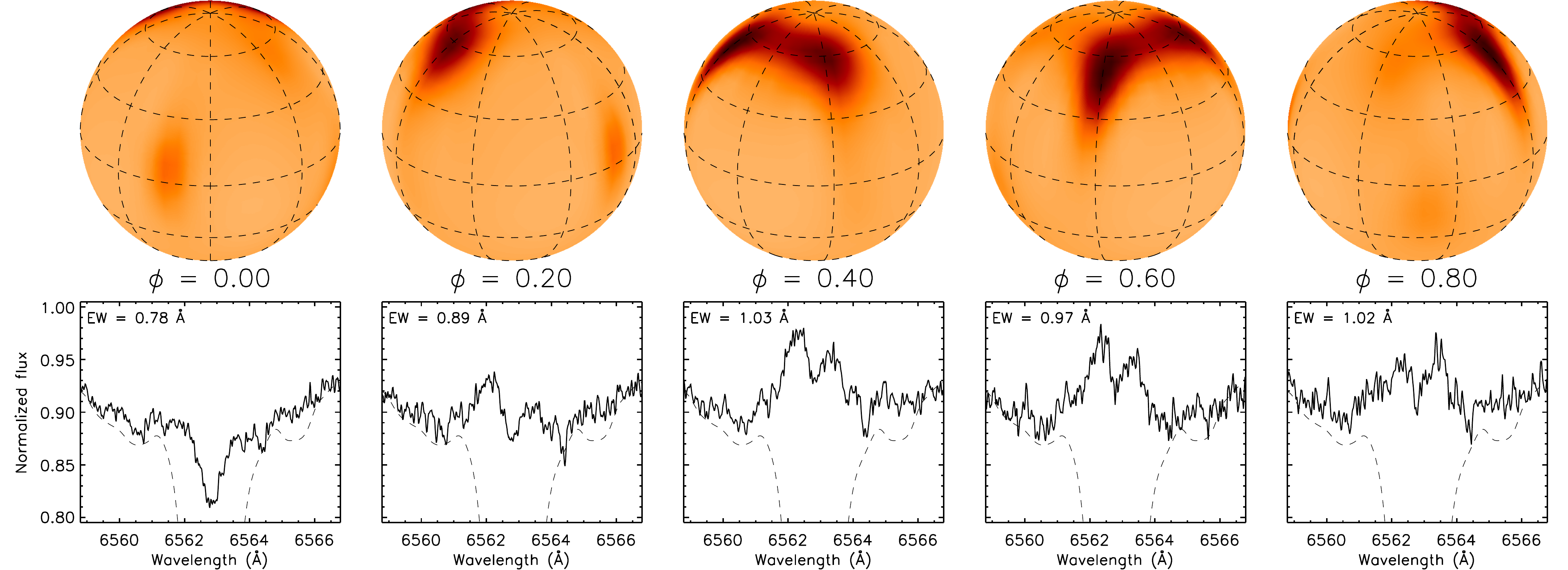}

\caption{Orthographic projection of the first map (data group 1 in Table \ref{tab:Time-intervals})
at five different rotational phases together with the average H$\alpha$
line at each phase. The spectrum of the non-active reference star
HD\,3765 is shown as a dashed line. The color table is the same as
in Fig.~\ref{fig:Maps}. \label{fig:HaMap1}}
\end{figure*}

\begin{figure*}[!t]
\includegraphics[width=1\textwidth]{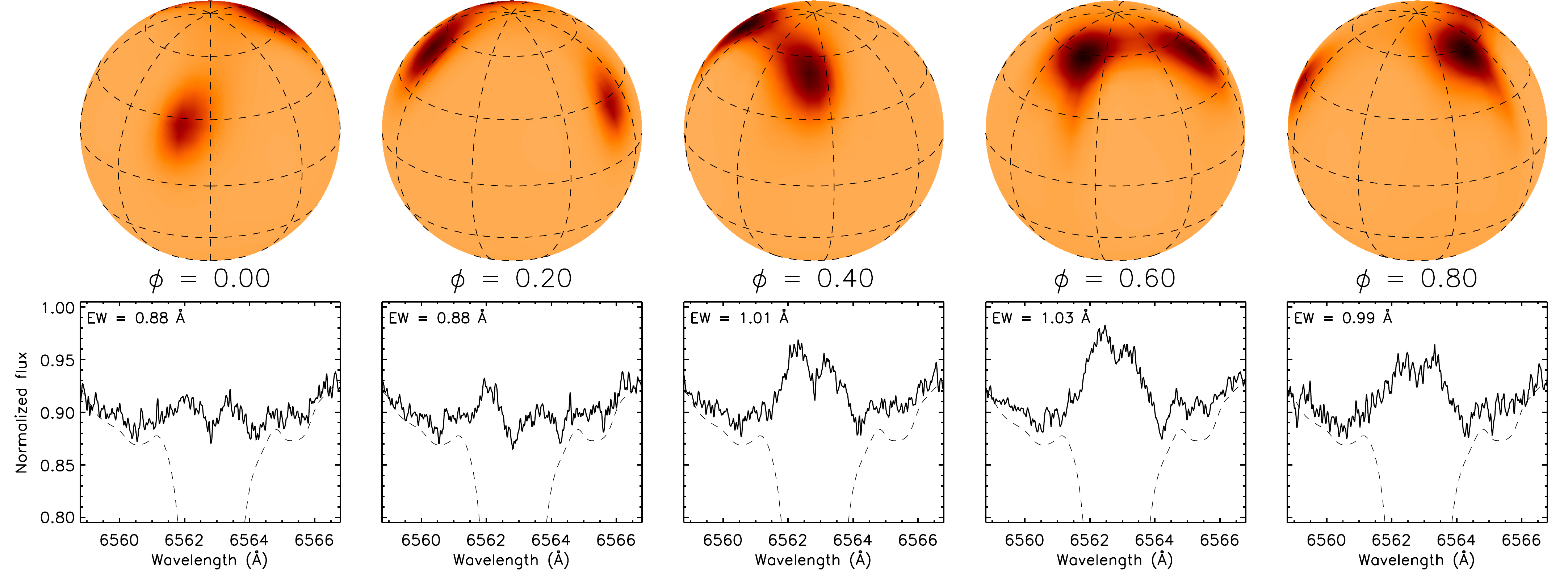}

\caption{Same as Fig. \ref{fig:HaMap1} but for the second map (data group
2 in Table \ref{tab:Time-intervals}).\label{fig:HaMap2}}
\end{figure*}

\begin{figure*}
\includegraphics[width=1\textwidth]{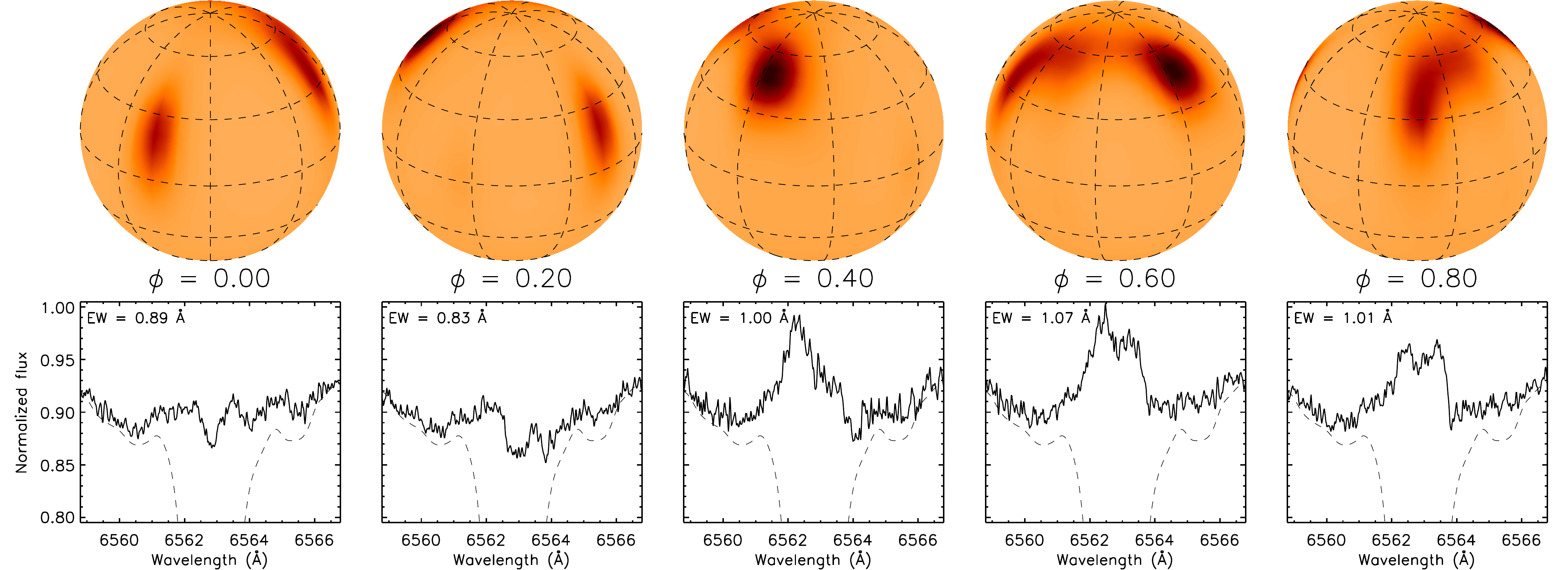}

\caption{Same as Fig. \ref{fig:HaMap1} but for the third map (data group 3
in Table \ref{tab:Time-intervals}). \label{fig:HaMap3}}
\end{figure*}

\begin{figure*}[!t]
\includegraphics[width=1\textwidth]{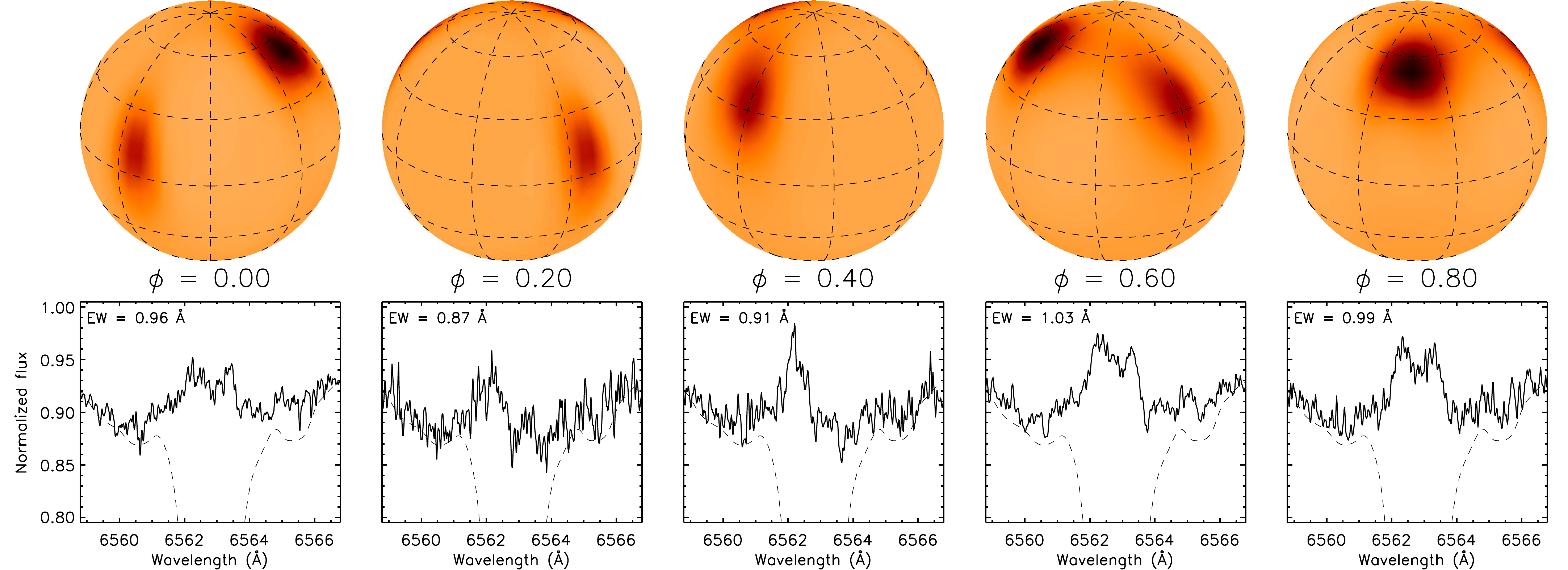}

\caption{Same as Fig. \ref{fig:HaMap1} but for the fourth map (data group
4 in Table \ref{tab:Time-intervals}). \label{fig:HaMap4}}
\end{figure*}

\begin{figure*}[t]
\includegraphics[width=1\textwidth]{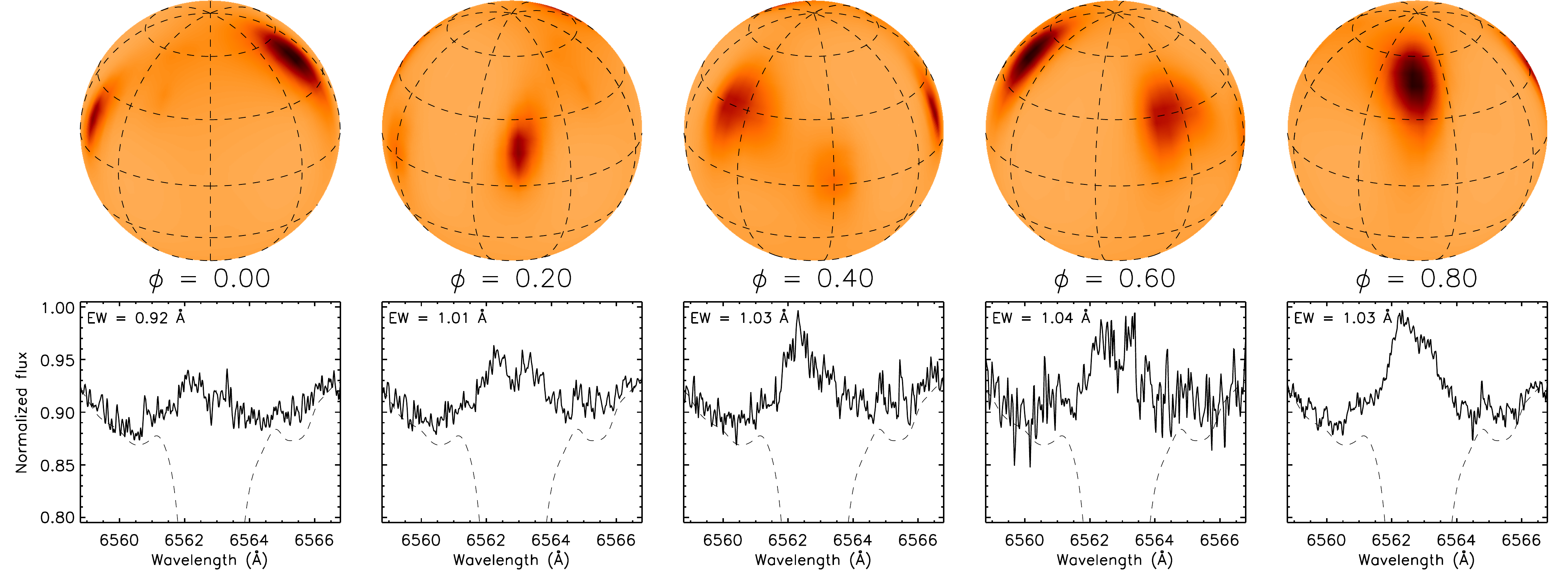}

\caption{Same as Fig. \ref{fig:HaMap1} but for the fifth map (data group 5
in Table \ref{tab:Time-intervals}). \label{fig:HaMap5}}
\end{figure*}

A way to assess the degree of spatial association between photospheric
and chromospheric active regions is by cross-correlating the photometric
light curve with the phase-resolved estimations of $\mathrm{EW_{\mathrm{H\alpha}}}$
and $\mathrm{EW_{\mathrm{H\beta}}}$ excluding flares. As this analysis
requires a very dense phase-coverage, the results from the first and
second groups of Table \ref{tab:Time-intervals} were combined to
increase the amount of photometric data-points, and the results from
the last two groups to increase the amount of chromospheric measurements.
Each one of the three sets was fit to a high-degree polynomial and
then cross-correlated. The result from the first set shows a lag of
$-0.05$ phases, where the negative value indicates that the photometry
is ahead in phase. Although this is not a negligible difference and
is observed for both H$\alpha$ and H$\beta$, it could be probably
just a consequence of the lower quality of the photometric measurements
during that period of time. The cross-correlation of the two other
data sets shows, on the other hand, differences between $-0.01$ and
0.02 phases, indicating a good match and no clear trend in the lag
either for H$\alpha$ or  H$\beta$. In comparison, the cross-correlation
between the chromospheric data from the first and last sets gives
a difference of 0.10 phases.

To study in more detail the evolution of the chromospheric activity
and its relation with the spots, we computed for each Doppler map
the average H$\alpha$ line at five equidistant rotational phases
(0.0, 0.2, 0.4, 0.6 and 0.8). Each average line was calculated using
all spectra within a window of $\pm0.1$ phases with equal weight.
H$\alpha$ lines with $EW\geq1.2$\,\AA\ and those affected by the
possible mass ejection were discarded. The resultant average H$\alpha$
lines are plotted along with the projected stellar hemispheres in
Figures \ref{fig:HaMap1} to \ref{fig:HaMap5}. 

We observe that the hemisphere centered around phase 0.0 increases
its average $\mathrm{EW_{\mathrm{H\alpha}}}$ emission by up to $\approx0.2$\,\AA\
as the spot S3 becomes visible at the receding side of the star. The
hemisphere centered at phase 0.2 corresponds to the region between
S1 and S2. Its average H$\alpha$ line shows a weak decay as S2 becomes
less visible (Fig. \ref{fig:HaMap1} to Fig. \ref{fig:HaMap3}), but
increases again by almost 0.1\,\AA\ when S1 reaches the central
meridian. We note that although the first map at phase 0.0 and the
last map at phase 0.2 are both centered near S1, their respective
H$\alpha$ lines are significantly different, indicating that the
chromosphere above S1 became more active. The emission of the hemisphere
centered at phase 0.4 decays as S2 and S3 become less visible, but
recovers its strength when S1 appears. The $\mathrm{EW_{\mathrm{H\alpha}}}$
at phase 0.6 increases by 0.1\,\AA\ from the first to the third
map, when the view is centered at S3 as well as at the midpoint between
S2 and S3. No significant H$\alpha$ variations are observed around
phase 0.8, although the line in the first map could have been affected
by a minor flare.

\begin{figure*}
\includegraphics[width=1\textwidth]{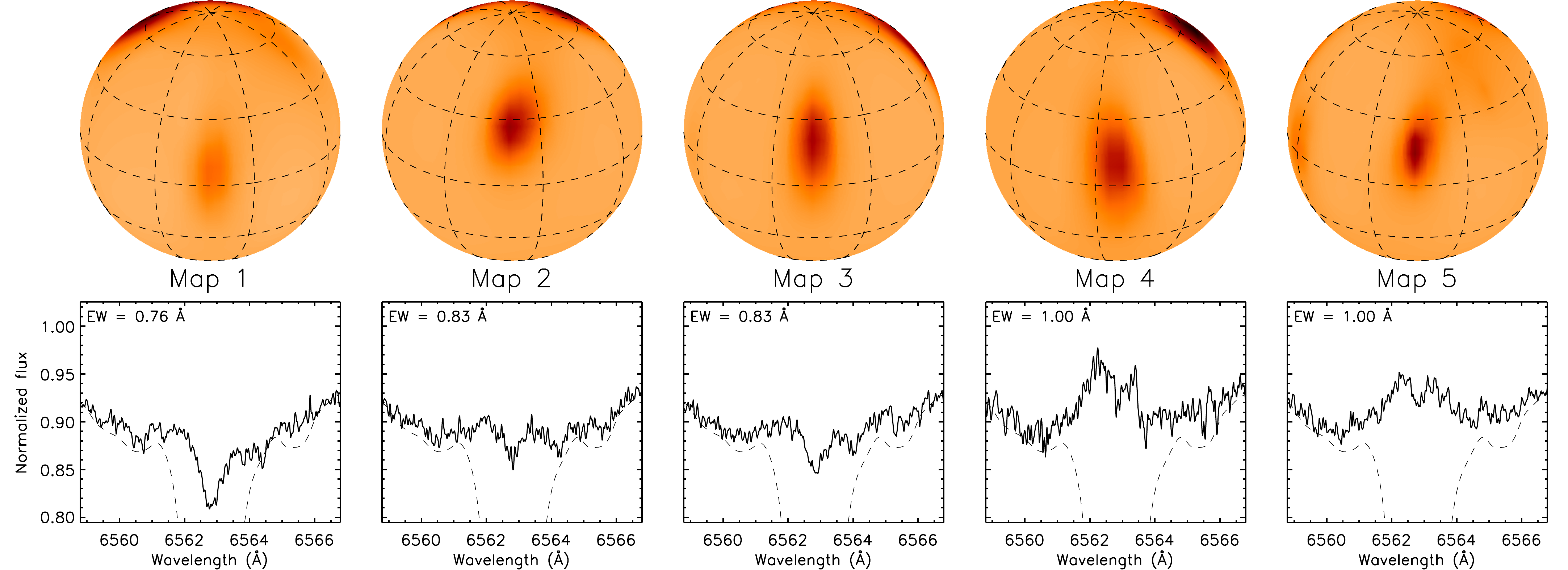}

\caption{Orthographic projection of the five maps centered at the position
of the spot S1 together with the average H$\alpha$ line at that phase.
The spectrum of the non-active reference star HD\,3765 is shown as
a dashed line. The color table is the same as in Fig.~\ref{fig:Maps}.
\label{fig:HaS1}}
\end{figure*}

\begin{figure*}
\includegraphics[width=1\textwidth]{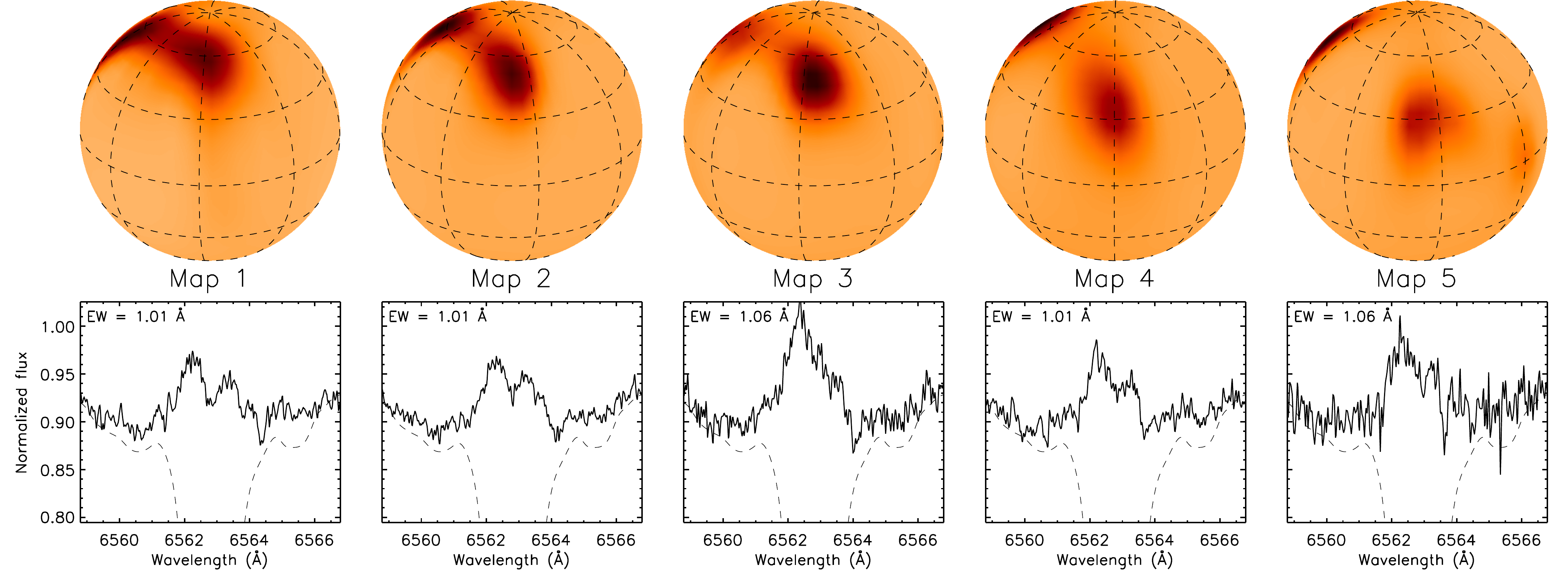}

\caption{Same as Fig. \ref{fig:HaS1} but for the spot S2.\label{fig:HaS2}}
\end{figure*}

\begin{figure*}
\includegraphics[width=1\textwidth]{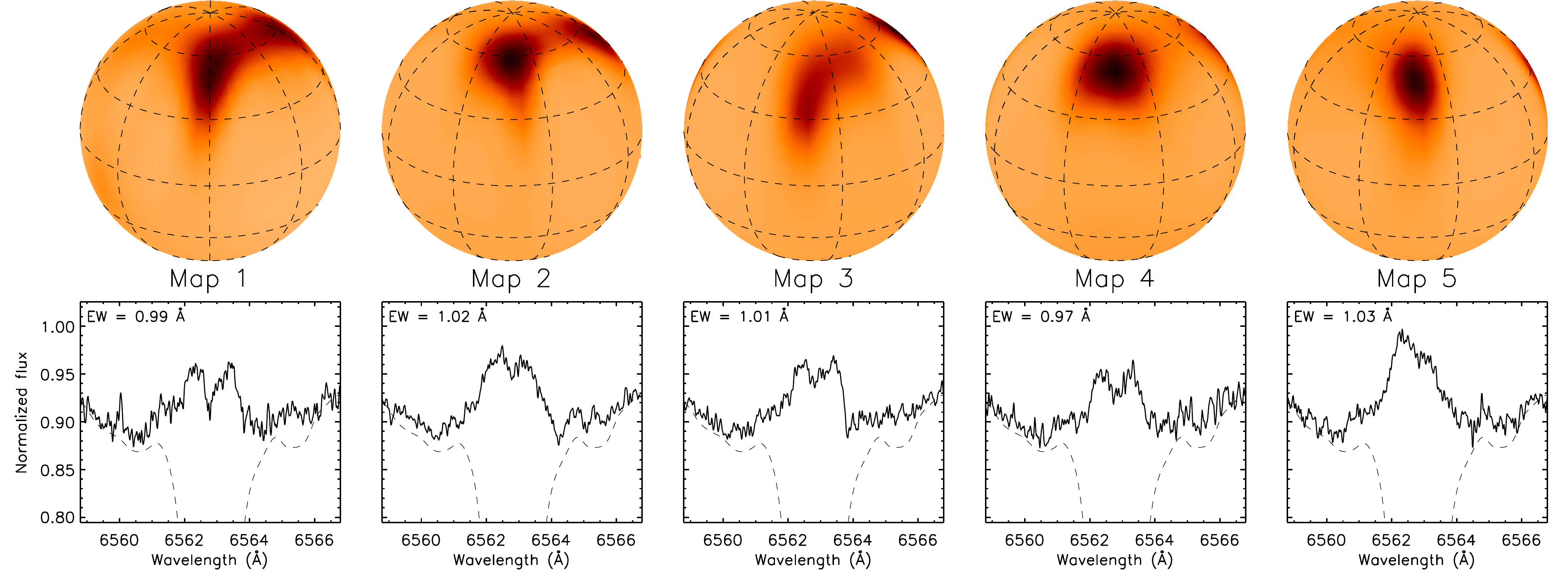}

\caption{Same as Fig. \ref{fig:HaS1} but for the spot S3.\label{fig:HaS3}}
\end{figure*}

\begin{figure*}
\includegraphics[width=1\textwidth]{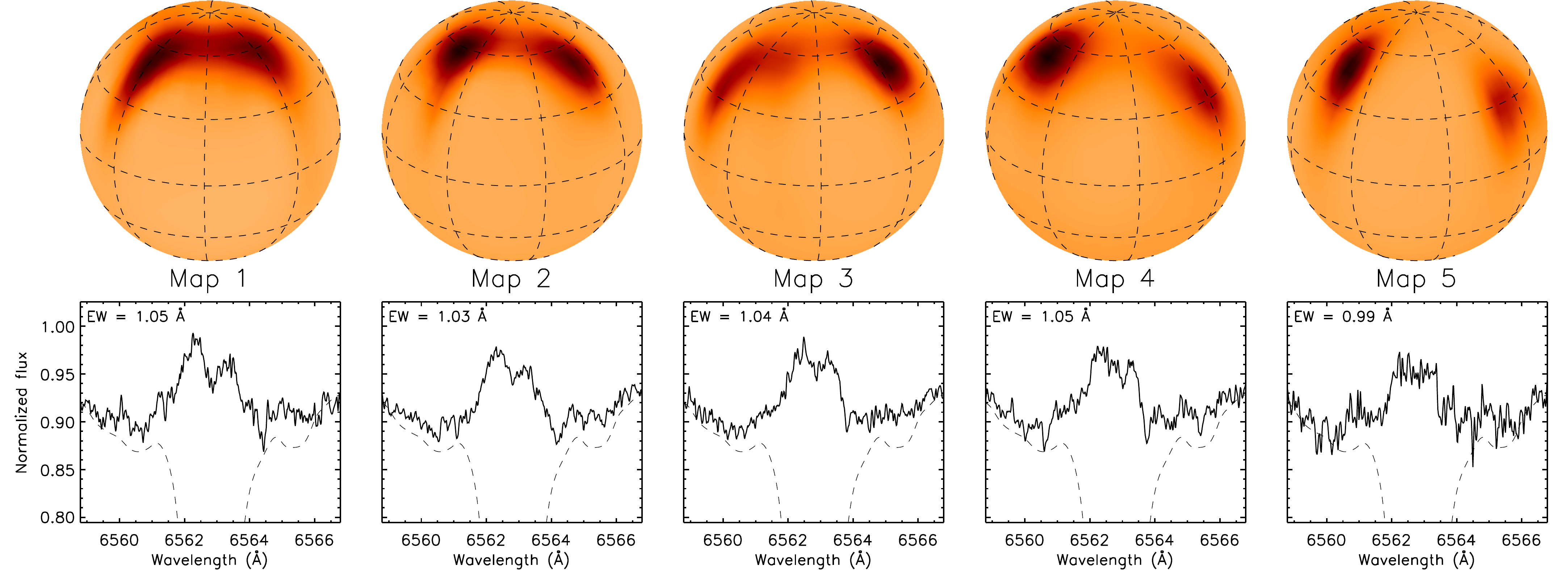}

\caption{Same as Fig. \ref{fig:HaS1} but for the midpoint between spots S2
and S3.\label{fig:HaS23}}
\end{figure*}

This analysis indicates good agreement between the spot locations
in the photospheric maps and the variations of the chromospheric H$\alpha$
line, but also that at least the chromospheric emission associated
with the spot S1 evolved significantly. To study the evolution of
the chromospheric activity of each active region we repeated the same
analysis as before but centering the H$\alpha$ profiles at the phases
of the three spots. In addition, we also included in this analysis
the region between the spots S2 and S3 to see whether the chromospheric
emission changes as they move apart from each other. The result can
be seen in Figures \ref{fig:HaS1} to \ref{fig:HaS23}. Although the
contribution of each active region can not be easily isolated, it
confirms that S1 is the only spot with a significant variation in
its chromospheric activity. On that side of the star the EW of the
average H$\alpha$ line increases from 0.76\,\AA\ to 1.00\,\AA.
Spots S2 and S3, on the other hand, only show fluctuation of 0.05\,\AA.
Similarly, the region between S2 and S3 only shows a minor decay in
the H$\alpha$ line of the last map. 

The structural changes of the H$\alpha$ line as the star rotates
can also provide some valuable information. The line usually presents
a central self reversal and thus two peaks, one on each side of the
central wavelength. The two peaks have a similar strength when the
active regions are near the meridian. Nevertheless, when one of the
spots appears at the limb moving toward the central meridian, the
blue peak can be up to a 10\% stronger than the other in terms of
peak intensity. This effect is particularly evident when the receding
side of the star is unspotted (see, e.g., phase 0.2 in Fig.~\ref{fig:HaMap1}
and phase 0.4 in Fig.~\ref{fig:HaMap4}). On the other hand, H$\alpha$
lines observed when the active region is on the receding side show
either two peaks of similar intensity or also the blue peak stronger
than the red (see, e.g., phase 0.2 Fig.~\ref{fig:HaMap4}). What
we never observed is an asymmetry dominated by the red peak. This
indicates that the rotational Doppler shift can be discarded as the
sole reason of the line asymmetries.

The region near the red peak of the H$\alpha$ line is affected by
several telluric lines. To test whether the tellurics are the reason
why we never observed the H$\alpha$ profile dominated by a red peak,
we simulated their impact in a worst case scenario. Because of the
intrinsic variability of the H$\alpha$ line and the limited S/N,
we isolated the telluric contribution using STELLA spectra of the
telluric standard star HD~177724. These spectra were taken when the
SES spectrograph was already operating in its nominal resolving power
of $R=55\,000$. To enable the direct comparison with the LQ~Hya
spectra, we reduced the resolution of the HD~177724 spectra to $R=30\,000$
with a Gaussian convolution. We then compared both sets of data and
chose an HD~177724 spectrum with the telluric water vapor lines slightly
stronger than in the most affected LQ~Hya spectrum.

The only telluric line that could have a significant impact on the
red peak of the H$\alpha$ line moves monotonically in the LQ~Hya
spectra from $\lambda_{\mathrm{ini}}=6564.5$\,\AA\ at the beginning
of the observations to $\lambda_{\mathrm{end}}=6563.4$\,\AA\ at
the end. This implies that the telluric line is near the center of
the red peak only at the end of the observations. In the data from
Figs.~\ref{fig:HaMap1} to \ref{fig:HaMap4} the telluric line is
too far away to suppress the red peak. Additionally, we also reversed
H$\alpha$ profiles dominated by a blue peak around the wavelength
$\lambda_{\mathrm{H\alpha}}=6562.79$\,\AA. In this way, the blue
peak appears in the position of the red peak. We used then the telluric
template to contaminate those profiles. We see that even in the worst
case scenario the telluric line is too sharp compared to the red peak
to completely suppress it. An example is shown in the appendix in
Fig.~\ref{fig:Tellu_en_Ha}. We can therefore conclude that the telluric
lines can explain second order asymmetries but not the fact that some
H$\alpha$ lines have only a blue peak, nor that the red peak never
dominates the profile. 

Considering that the blue asymmetry is strongest when the spot is
on the limb, and that it disappears when the spot is on the meridian,
we suggest that the source could be a flow of moving magnetic features
surrounding the spots, parallel to the stellar surface, and maybe
analogous to the solar Evershed effect or to the moat flow.

More difficult to explain in view of our data is the sharp decay of
the sporadic flare activity. These flares were habitual during the
first 20 days of observations but then they became more occasional
and delocalized from the most active hemisphere. This corresponds
to the epoch of closest distance between the spots S2 and S3, suggesting
that the flares could have been the result of the interaction between
the magnetic fields of these two active regions. Nonetheless, data
from the second group of Table \ref{tab:Time-intervals} (days 22-53)
show an increment in the separation between the spots of only 0.02
phases and practically no sporadic flares. This change happened without
a clear counterpart neither in the photometry, nor in the basal chromospheric
emission. 

\section{Summary and conclusions\label{sec:Summary-and-conclusions}}

We have analyzed four months of quasi-simultaneous spectroscopic and
photometric observations of the young K2 dwarf LQ~Hya to study the
connection between photospheric spots and chromospheric active regions,
as well as the evolution of the activity in the short term.

The Doppler maps show that the stellar surface was dominated by three
spots; a low-latitude spot (S1) and two bigger spots at higher latitudes
(S2 and S3). The three spots are present in all maps, indicating that
the active regions of LQ~Hya can live for at least four month. The
combined use of the \emph{V}-band photometry and the Doppler maps
reveal that the changes in the photosphere of the star were mostly
a consequence of the spatial redistribution of the spots. The photometric
light curve presents clear rotational modulation with minimum brightness
at the midpoint between S2 and S3. As the spots moved apart from each
other the amplitude of the modulation decreased from 0\fm16 at the
middle of the observations to 0\fm09 at the end. These process took
place while the mean brightness of the star remained constant at a
value of $V\approx7.90$\,mag.

The chromospheric emission shows a good anti-correlation with the
photometric light curve, but also some differences. The side of the
star with highest emission was centered at the midpoint between the
spots S2 and S3, and moved toward higher phases with them. Curiously,
the basal emission from that side of the star remained constant while
S2 and S3 move apart from each other instead of declining in anti-correlation
with the brightness. The study of the H$\alpha$ and H$\beta$ lines
as the star rotates reveals the coexistence of chromospheric active
regions and photospheric dark spots. The chromospheric emission associated
to the spots S2 and S3 remained roughly constant but increased for
the spots S1. This led to an overall increment of the chromospheric
basal emission that seems to be somewhat disconnected from the photosphere.
Additionally, the asymmetries at the core of the H$\alpha$ line suggest
the existence of a radial outflow surrounding the spots.

We also detected higher flare activity during the first twenty days
of observations. One of the flares produced an enhancement at the
blue wing of the H$\alpha$ line that lasted for more that two days.
From the shape of the velocity profile and the lack of rotational
modulation, we attributed the effect to a mass ejection.
\begin{acknowledgements}
We are grateful to the State of Brandenburg and the German federal
ministry for education and research (BMBF) for their continuous support
of the STELLA and APT activities. The STELLA facility is a collaboration
of the AIP in Brandenburg with the IAC in Tenerife. We thank Lou Boyd
for nursing the APTs at Fairborn Observatory, and Thomas Granzer for
his help with the photometric data. Finally, we would also like to
thank our anonymous referee for the helpful comments and suggestions.
\end{acknowledgements}

\bibliographystyle{aa}
\bibliography{biblio_EvoLQHya}

\clearpage

\appendix

\section{Example H$\alpha$ lines with broad wings and relatively low emission\label{sec:Apendice_EjemHa}}

\begin{figure}
\includegraphics[width=0.9\columnwidth]{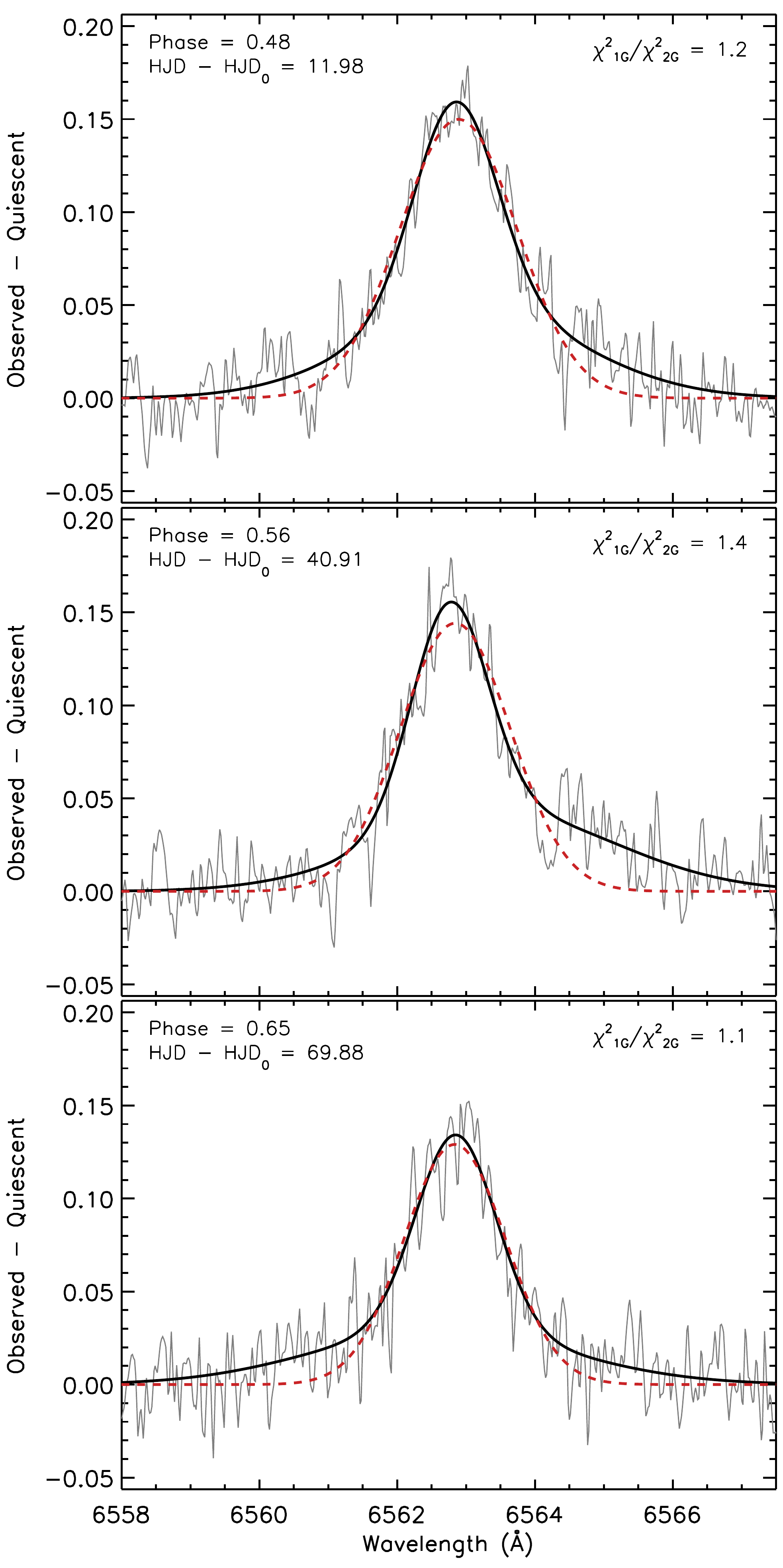}

\caption{Example of three subtracted H$\alpha$ lines showing broad wings despite
their relatively low emission. The EW of the net emission profiles
is between 1.0 and 1.1\,\AA. The subtracted profiles and their corresponding
two- and single-Gaussians fits are shown as a gray thin, solid black,
and dashed red line, respectively. The ratio of the chi-square goodness
of the fits is indicated on the right side of each panel. \label{fig:Example-of-three}}
\end{figure}

Roughly a half of the H$\alpha$ lines from the basal emission with
EWs in the range 1.0-1.1\,\AA\ show broad wings that could be a
consequence of micro-flare activity. Fig.~\ref{fig:Example-of-three}
shows a few examples.

\section{Observed and inverted line profiles.}

\begin{figure*}
\includegraphics[width=1\textwidth]{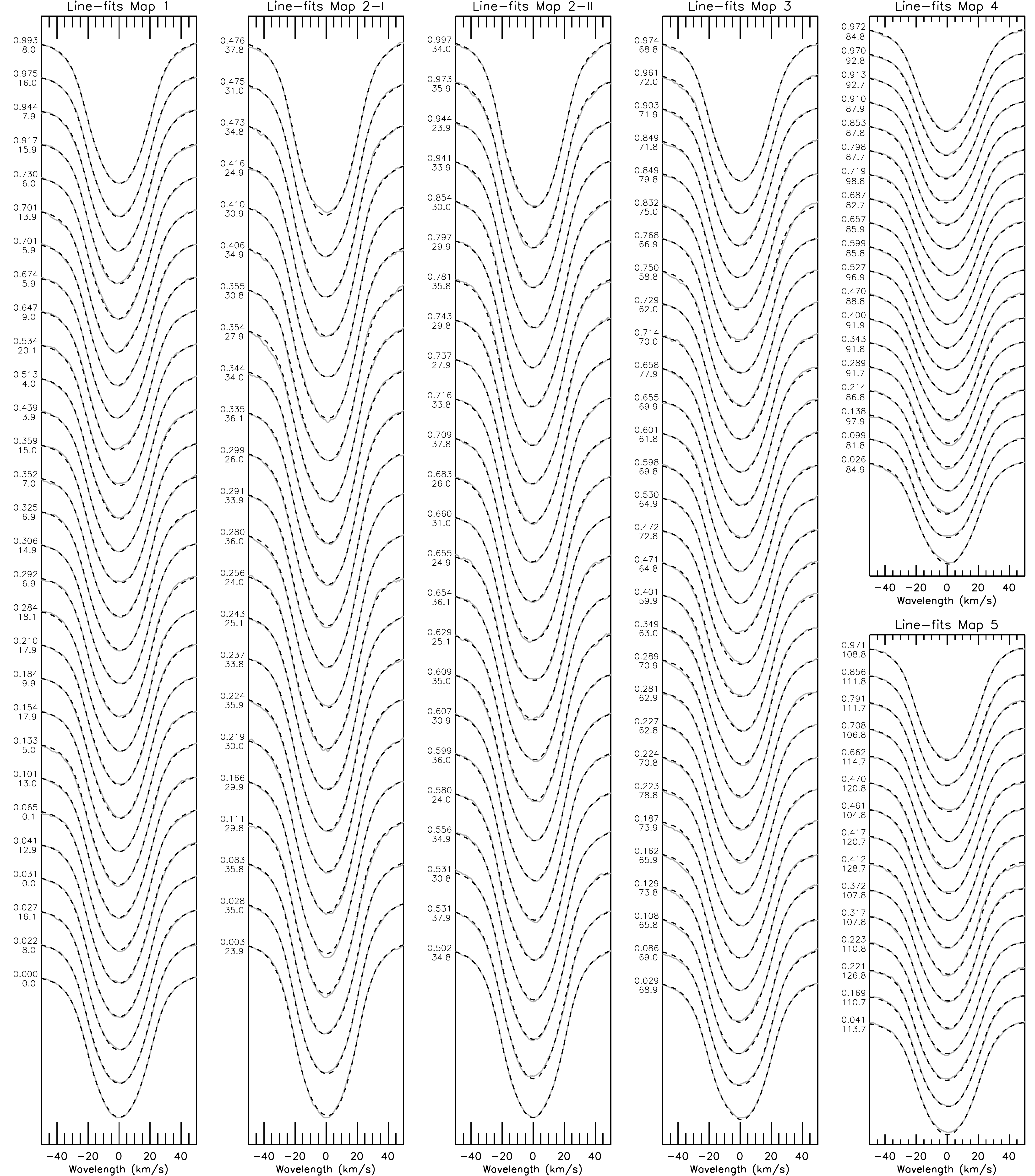}

\caption{Average line profiles used for the inversion (gray solid) together
with their model fit (black dashed). The two numbers on the left side
of each line indicate the rotational phase (top) and the relative
time $\mathrm{HJD}-\mathrm{HJD_{0}}$ (bottom). \label{fig:Line-fits}}
\end{figure*}

Figure~\ref{fig:Line-fits} shows the average profiles of the 19
individual lines used for Doppler imaging, together with the averaged
inverted lines. The rotational phase and the date of each spectrum
are also indicated.

\section{Example of map affected by distorted lines}

\begin{figure}[H]
\includegraphics[width=0.9\columnwidth]{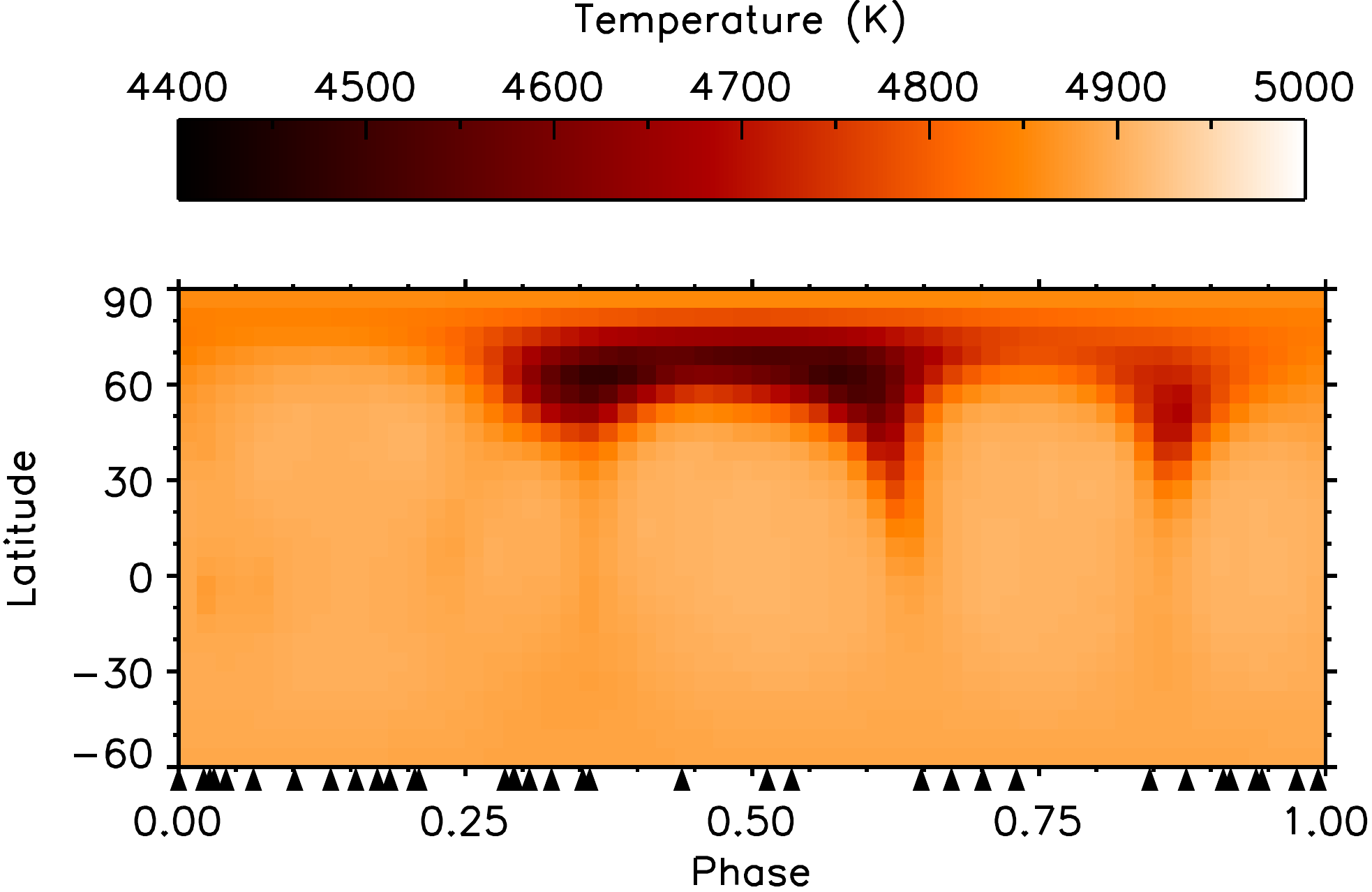}

\caption{Doppler map computed in the same manner as the first map in Fig.~\ref{fig:Maps},
but including seven spectra affected by a possible mass ejection from
a flare. As a result of this, the near equatorial spot from phase
0.1 disappears, the resolution at high latitudes get worse, and an
artifact near phase 0.85 appears. \label{fig:Map_G1SpcCME}}
\end{figure}

Spectral lines affected by high levels of noise, or distorted by activity
phenomena other than spots, can produce spurious features in the Doppler
maps. In Fig.~\ref{fig:Map_G1SpcCME} we show as an example the effect
that seven spectra affected by a possible mass ejection have on the
first map of Fig.~\ref{fig:Maps}.

\section{Example of the effect of telluric lines on the red peak of the H$\alpha$
line}

\begin{figure}[H]
\includegraphics[width=0.9\columnwidth]{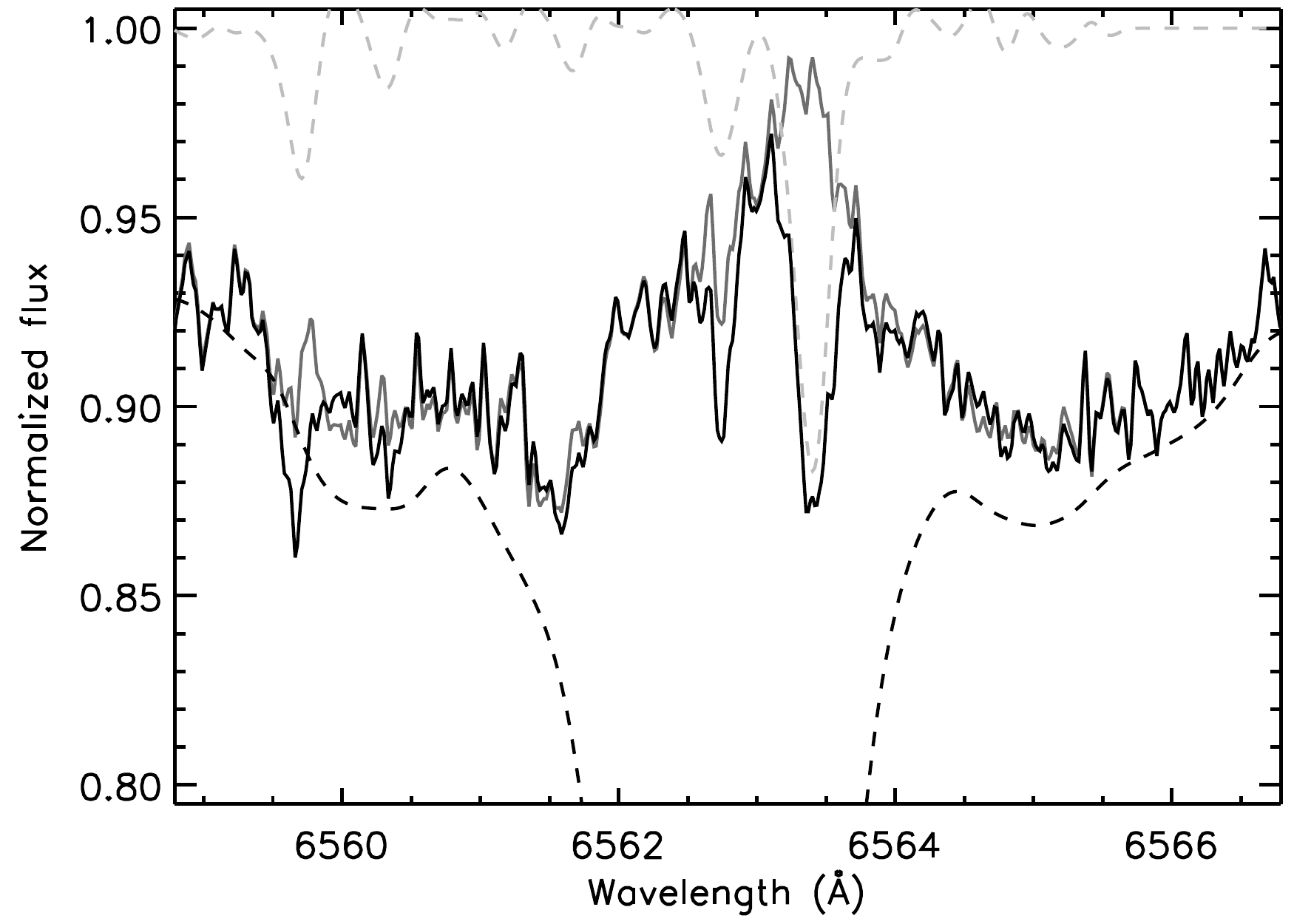}

\caption{Simulated effect of telluric lines on the red peak of the H$\alpha$
line in a worst case scenario. The black-solid line is the result
of multiplying the gray-solid line with the telluric template (gray
dashed). The non-active reference spectrum of HD\,3765 is shown as
a black-dashed line. \label{fig:Tellu_en_Ha}}

\end{figure}

In Fig. \ref{fig:Tellu_en_Ha} we simulate an H$\alpha$ profile dominated
by a red peak reversing the H$\alpha$ line of Fig. \ref{fig:HaMap3}
and phase 0.4. To simulate the effect of telluric lines in a worst
case scenario we chose a template with telluric lines slightly stronger
that in any LQ~Hya spectrum. The template was shifted to mach the
closest distance between the strongest telluric line and the red peak
compatible with the observations.

\end{document}